\documentclass[authoryear,preprint,review,12pt,a4paper]{elsarticle}

\usepackage{amsmath,amssymb}
\usepackage[nodots]{numcompress}
\usepackage{supertabular,multirow}
\usepackage{graphicx}
\usepackage{epsfig}
\usepackage{subfigure}
\usepackage{eufrak}
\usepackage{bm}
\usepackage{color}
\usepackage{algpseudocode}
\usepackage{algorithm}

\def\ie{{i.e.\ }}

\journal{Computer Methods in Applied Mechanics and Engineering}

\begin{document}

\begin{frontmatter}

\title{Development of $hp$-inverse model by using generalized polynomial chaos} 

\author[IBM]{Kyongmin Yeo\corref{cor1}}
\author[addr1]{Youngdeok Hwang}
\author[addr2]{Xiao Liu}
\author[IBM]{Jayant Kalagnanam}

\cortext[cor1]{kyeo@us.ibm.com}
\address[IBM]{IBM T.J. Watson Research Center, Yorktown Heights, NY 10598, USA}
\address[addr1]{Department of Statistics, Sungkyunkwan University, Seoul, Korea}
\address[addr2]{Department of Industrial Engineering, University of Arkansas, Fayetteville, AR 72710, USA}

%
%


	
	\begin{abstract}
	We present a $hp$-inverse model to estimate a smooth, non-negative source function from a limited number of observations for a two-dimensional linear source inversion problem. A standard least-square inverse model is formulated by using a set of Gaussian radial basis functions (GRBF) on a rectangular mesh system with a uniform grid space. Here, the choice of the mesh system is modeled as a random variable and the generalized polynomial chaos (gPC) expansion is used to represent the random mesh system. It is shown that the convolution of gPC and GRBF provides hierarchical basis functions for the linear source inverse model with the $hp$-refinement capability. We propose a mixed $l_1$  and $l_2$ regularization to exploit the hierarchical nature of the basis functions to find a sparse solution. The $hp$-inverse model has an advantage over the standard least-square inverse model when the number of data is limited. It is shown that the $hp$-inverse model provides a good estimate of the source function even when the number of unknown parameters ($m$) is much larger the number of data ($n$), e.g., $m/n > 40$.
	\end{abstract}
	
	
\end{frontmatter}
	
	\section{Introduction}
	
Air pollution, generated by either anthropogenic or natural causes, poses a major public health threat. Not only long-term \citep{Hoek13}, but also acute exposure \citep{Phalen11} over a certain threshold can cause health problems. Due to its immense importance, there have been substantial development in the computational modeling of the transport of air-borne pollution over the past decade \citep{Byun06,El-Harbawl13,Fast06}. However, prediction of air pollution by using these computational models requires extensive prior information on the distribution and magnitudes of pollution emission sources, which in most cases is incomplete or has high uncertainty \citep{Thunis16}. Moreover, in many cases, it is of greater interest to identify the source of pollution when abnormally high pollution is observed in the air quality monitoring network to mitigate a possible public health hazard. This atmospheric inverse problem  to find the pollution emission source using a set of measurements from a sensor network, has attracted significant attention in the atmospheric science community.
	
One of the fundamental building blocks of the inverse model is the atmospheric dispersion process, which is modeled by an advection-diffusion equation \citep{Stockie11}. 
Deterministic approaches, adopted from the field of atmospheric data assimilation, have been used widely for the source inverse problem \citep{Eckhardt08,Issartel07,Camara14,Pudykiewicz98}. In the deterministic approaches, typically a partial-differential-equation constrained optimization problem is solved to minimize a convex loss function, \emph{e.g.}, $l_2$-distance between the computational model prediction and the observations. Since the optimization formulation for an inverse model usually leads to underdetermined or ill-conditioned system, much research effort is focused on regularizing the solution. Recently, inverse models exploiting Bayesian inference have become popular \citep{Chow08,Keats07,Rajaona15}, due to the strength of the Bayesian methods in dealing with noisy and incomplete data.
In \citet{Keats07}, the adjoint advection-diffusion operator is used to reduce the computational cost. \citet{Hwang19} proposed an efficient Bayesian source inversion model to estimate the two-dimensional source function by exploiting the adjoint advection-diffusion operator. More general approaches to mitigate the high computational cost have been proposed by either accelerating the convergence of a Monte Carlo simulation \citep{Marzouk07}, constructing surrogate models \citep{Li14}, or developing a low-dimensional representation \citep{Lieberman10,Roosta14}.
	
Most of the previous inverse models consider either estimating the magnitudes of the source at each computational grid points by combining a large volume of heterogenous data \citep{deFoy15,Hwang17,Issartel07}, or finding the locations and magnitudes of one or a few point sources from a limited number of data \citep{Keats07,Marzouk07}. In this paper, we propose an inverse model based on a regularized optimization formulation to estimate a smooth source function from a small number of observations. First, we follow the conventional approach of approximating a smooth function by a set of Gaussian radial basis functions centered at the collocation points of a rectangular mesh system. Obviously, the solution of the inverse models is strongly dependent on the choice of the mesh system. To relax the dependency on the mesh system, we introduce a random mesh system, in which the choice of the mesh system is modeled as a random variable. A stochastic inverse model is formulated on this random mesh system and the generalized Polynomial Chaos expansion (gPC) \citep{Xiu07} is employed to tackle the stochastic inverse problem. It is shown that the stochastic formulation leads to a $hp$-inverse model, in which the unknown smooth function is approximated by hierarchical basis functions. The $hp$-inverse model has an advantage over the standard least-square inverse model, particularly when the number of data is limited, due to its capability of $hp$-refinement \citep{Karniadakis05}.
	
	
	
	This paper is organized as follows. Section \ref{sec:dispersion} describes a least-square formulation of the advection-diffusion problem by using an adjoint operator. In section \ref{sec:uncertainty}, we reformulate the deterministic least-square problem as a stochastic problem by using gPC. In section \ref{sec:optimization}, a mixed $l_1$- and $l_2$-regularization is introduced to exploit the hierarchical nature of the basis functions and an algorithm based on the alternating direction method of multipliers is presented to solve the optimization problem. The proposed inverse model is tested in section \ref{sec:results}. Finally, the concluding remarks are given in section \ref{sec:summary}.
	
\section{Least-square inverse model}\label{sec:dispersion}
	
\subsection{Forward model}\label{sec:forward}
	
	We consider the following advection-diffusion problem,
	\begin{equation}\label{eqn:adv-diff}
	\begin{cases}
	\mathcal{A} \phi(\bm{x},t) = Q(\bm{x})& \bm{x} \in D\\
	\bm{n \cdot \nabla} \phi(\bm{x},t) = 0 & \bm{x} \in \partial D_{out}\\
	\phi(\bm{x},t) = 0 & \bm{x} \in \partial D_{in}\\
	\end{cases}.
	\end{equation}
	Here, $D$ is a rectangular domain $D = (X_1,X_1+L_1) \times (X_2,X_2+L_2)$, in which $(X_1,X_2)$ is the coordinate of the lower left corner of the domain and $L_1$ and $L_2$ are the lengths in the $x_1$ and $x_2$ directions, respectively, with the boundary $\partial D$, and $\bm{n}$ denotes an outward normal vector on $\partial D = \partial D_{out} \cup \partial D_{in}$. 
The outflow and inflow boundaries are defined in terms of the fluid velocity $\bm{u}(\bm{x},t)$ as $\partial D_{out} = \{\bm{x}:\bm{x} \in \partial D,~\bm{n \cdot u}(\bm{x},t) \ge 0\}$ and $\partial D_{in} = \{\bm{x}:\bm{x} \in \partial D,~\bm{n \cdot u}(\bm{x},t) < 0\}$. The fluid velocity is assumed to be given by a measurement or a computational fluid dynamics model. The advection-diffusion operator is defined as
	\begin{equation} \label{eqn:governing}
	\mathcal{A} \phi(\bm{x},t) = \frac{\partial}{\partial t}\phi(\bm{x},t) + \frac{\partial}{\partial x_j} (u_j(\bm{x},t)\phi(\bm{x},t))-\frac{\partial}{\partial x_i} \left( K_{ij}(\bm{x},t) \frac{\partial}{\partial x_j} \phi(\bm{x},t) \right),
	\end{equation}
	in which $\bm{K}(\bm{x},t)$ is a symmetric second-order tensor of the diffusivity. We assume that the elements of $\bm{u}(\bm{x},t)$ and $\bm{K}(\bm{x},t)$ are smooth functions with uniformly bounded derivatives of all orders. The source strength $Q(\bm{x})$ is the unknown function, but assumed to be smooth. Furthermore, we consider a non-negative source, i.e., $Q(\bm{x}) \ge 0$ for every $\bm{x} \in D$.  
	
	Contrary to the usual computational prediction problem, where equation (\ref{eqn:adv-diff}) is solved for known $Q(\bm{x})$ to estimate $\phi(\bm{x},t)$ at the sensor locations, the inverse model aims to estimate $Q(\bm{x})$ from the given observations $\Phi(t^o)$ at the observation time $t^o$. Here, instead of estimating $Q(\bm{x})$ directly by solving an infinite dimensional optimization problem, $Q(\bm{x})$ is approximated by the sum of a set of basis functions to reduce the dimensionality of the problem. 
	Let $Q^*(\bm{x})$ be a finite-dimensional approximation,
	\begin{equation}
	Q^*(\bm{x}) = \sum_{i=1}^{N_k} \beta_i \mathcal{P}_i(\bm{x}).
	\end{equation}
	Here, $\bm{\mathcal{P}}$ is a set of basis functions, $N_k$ is the total number of the basis functions and $\bm{\beta} = (\beta_1,\ldots, \beta_{N_k})$ denotes the coefficients. The problem of estimating a continuous surface is reduced to a problem of finding $N_k$ coefficients, $\bm{\beta}$. There are many possible choices for the basis functions as long as $Q^*(\bm{x})$ satisfies the non-negativity condition: $\sum \mathcal{P}_i(\bm{x}) \beta_i \ge 0$. 
	Here, a set of Gaussian radial basis functions (GRBF) located at the collocation points of a rectangular mesh is used as the basis functions;
	\begin{equation}
	\mathcal{P}_i(\bm{x}) = \frac{1}{2\pi (c \Delta)^2} \exp \left[ - \frac{1}{2} \frac{ |\bm{x}-\bm{y}^i|^2}{(c \Delta)^2} \right],
	\end{equation} 
	in which $\Delta$ is the distance between the neighboring collocation points, $\bm{y}^i$ is the location of $i$-th collocation point and $c$ is an $O(0.1)$ parameter. 

	\begin{figure}
		\centering
		\includegraphics[width=0.4\textwidth]{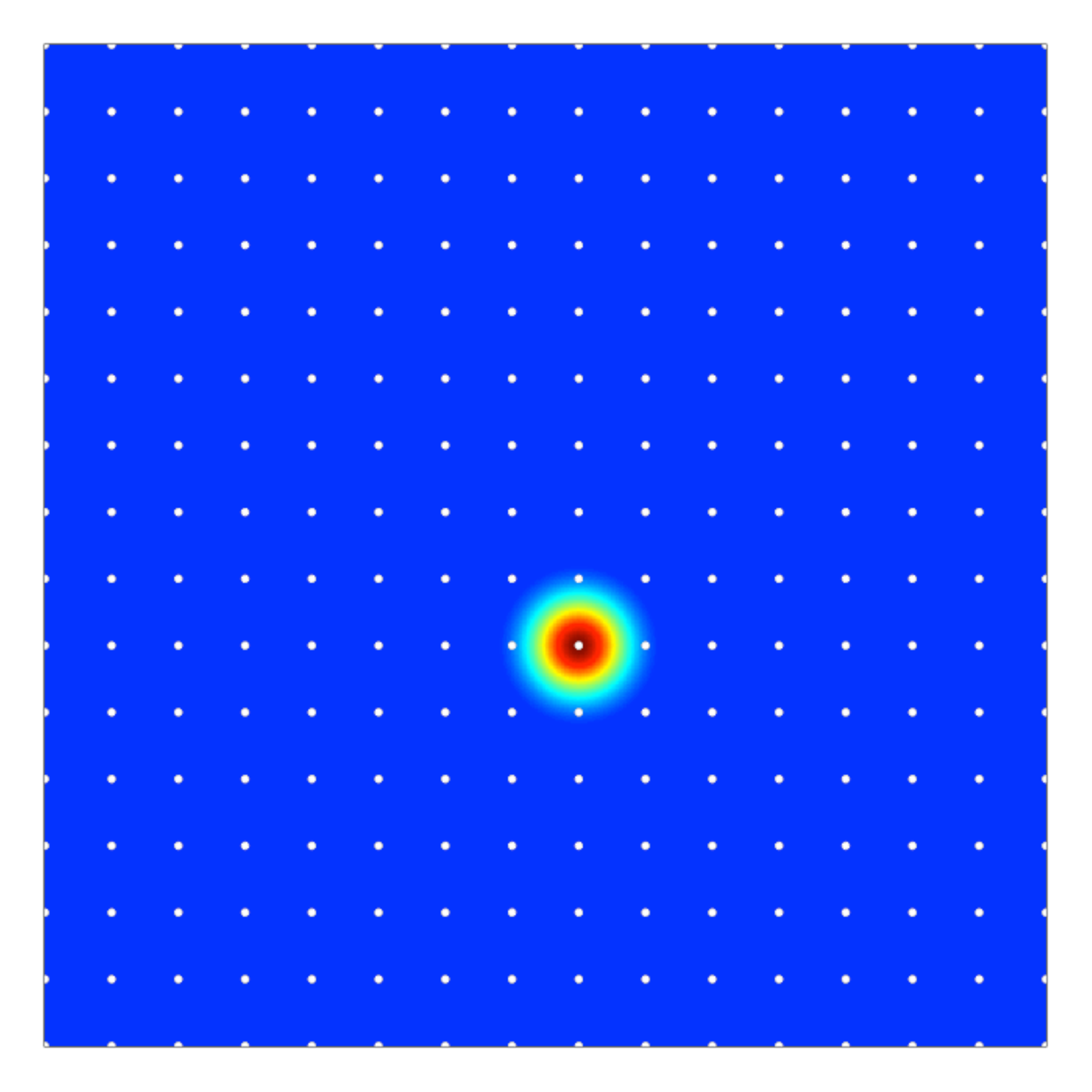}
		\caption{ Example of the collocation points (white dot) and a basis function $\mathcal{P}_i(\bm{x})$. }\label{fig:collocation}
	\end{figure}

The rectangular mesh system is defined by a tensor product of two one-dimensional collocation sets,
	\begin{equation}
	\mathcal{W} = \bm{s}_1 \otimes \bm{s}_2.
	\end{equation}
	Here, $\bm{s}_1$ and $\bm{s}_2$ denote the sets of collocation points in the $x_1$- and $x_2$-directions, respectively;
	\begin{align}
	\bm{s}_1 &= \{x_1^i; x^i_1 = x^0_1+i \times \Delta,~ i = 1,\cdots,N^1_k \}, \nonumber\\
	\bm{s}_2 &= \{x_2^i; x^i_2 = x^0_2+i \times \Delta,~ i = 1,\cdots,N^2_k \}, \nonumber
	\end{align}
	in which $x_1^0$ is the left and $x_2^0$ is the bottom end of the mesh system, and $N^j_k$ is the number of the collocation points in the $j$-direction, $N_k = N^1_k \times N^2_k$. An example of $\mathcal{W}$ and GRBF is shown in figure \ref{fig:collocation}. 
Obviously, in the limit of $\Delta \rightarrow 0$, $\mathcal{P}$ converges to the Dirac delta function $\mathcal{P}_i(\bm{x}) \rightarrow \delta (\bm{x}-\bm{y}^i)$ and, hence, $Q^*(\bm{x})$ converges uniformly to $Q(\bm{x})$. Using GRBF, the non-negativity condition can be satisfied by $\beta_i \ge 0$ for $i = 1,\cdots,N_k$. 
	
	Now, the advection-diffusion equation can be written as
	\begin{equation} \label{eqn:dispersion_operator}
	\mathcal{A}\phi(\bm{x},t) =  \sum_{i=1}^{N_k} \beta_i \mathcal{P}_i(\bm{x}).
	\end{equation}
	Since $\mathcal{A}$ is a linear operator, we can exploit the superposition of the solutions. Let $\phi_i(\bm{x},t;\beta_i)$ be the solution for the $i$-th GRBF;
	\[
	\mathcal{A}\phi_i(\bm{x},t;\beta_i) =  \beta_i \mathcal{P}_i(\bm{x}).
	\]
	Then, clearly,
	\[
	\phi(\bm{x},t) = \sum_{i=1}^{N_k} \phi_i(\bm{x},t;\beta_i).
	\]
	Moreover, from the linearity of $\mathcal{A}$,
	\begin{equation} \label{eqn:unit_strength}
	\phi_i(\bm{x},t;\beta_i) = \widetilde{\phi}_i(\bm{x},t)\beta_i.
	\end{equation}
	Here, $\widetilde{\phi}_i$ is the solution for $\mathcal{P}_i$ with a unit strength, \emph{i.e.}, $\beta_i =1$.
	
	The computational model output is related to the observation by an inner product with respect to a sensor function ($\chi$) as
	\begin{equation}\label{eqn:subset}
	\Phi_i(t^o) = \langle \phi(\bm{x},t),\chi(\bm{x},t;\bm{x}_i^o,t^o) \rangle + \epsilon_i ~~\text{for}~i = 1,\cdots,N_o,
	\end{equation}
	in which $\Phi_i(t^o)$ is the observation at the $i$-th sensor, $\bm{x}^o_i$ is the location of the  $i$-th sensor, $t^o$ is the time of the measurement, $N_o$ is the total number of the sensors, and $\epsilon$ is a Gaussian white noise representing the errors in the measurement as well as the computational model. 
	The angle bracket denotes an inner product
	\[
	\langle a(\bm{x},t),b(\bm{x},t) \rangle = \int_{\bm{x}\in D}\int^0_{-\infty} a(\bm{x},t+\tau)b(\bm{x},t+\tau) d\tau\,d\bm{x}.
	\]
	The sensor function depends on the types of the sensor or the data used in the analysis. 
	In this study, the sensor function is defined as
	\begin{equation}\label{eqn:sensor}
	\chi(\bm{x},t;\bm{x}_i^o,t^o) = \frac{1}{T_\chi} \delta(\bm{x}-\bm{x}_i^o) \{H(t - t^o + T_\chi) -H(t-t^o)\}.
	\end{equation}
	Here $H(t)$ is a Heaviside function, which is zero for $t < 0$ and one otherwise, and $T_\chi$ is an time-average window of the sensor.
	
	By comparing (\ref{eqn:unit_strength}) and (\ref{eqn:subset}), $\bm{\beta}$ can be related to the measurement by
	\begin{equation} \label{eqn:poll_approx}
	\bm{\Phi} = \widetilde{\bm{X}} \bm{\beta} + \bm{\epsilon}.
	\end{equation}
	Here, $\bm{\Phi}^T = (\Phi_1(t^o),\cdots,\Phi_{N_o}(t^o))^T$ and 
	\begin{equation}
	\widetilde{X}_{ij} = \langle \, \chi(\bm{x},t;\bm{x}_i^o,t^o),\widetilde{\phi}_j(\bm{x},t) \, \rangle.
	\end{equation}
	The dispersion matrix $\widetilde{\bm{X}} \in \mathbb{R}^{N_o\times N_k}$ relates $\bm{\beta}$ to the observation $\bm{\Phi}$.
	Since $\widetilde{\bm{X}}$ can be computed by solving exactly the same partial differential equation (\ref{eqn:governing}) for each GRBF ($\mathcal{P}_i$), the same numerical solver can be recycled.
	
	The source strength $Q^*(\bm{x})$ can be obtained by finding $\bm{\beta}$ from the following least-square minimization problem
	\begin{align} \label{eqn:optimization-1}
	\underset{{\bm{\beta}} \in \mathbb{R}^{N_k}, \bm{\beta}\geq 0 }{\text{arg min}} \frac{1}{2} \| \bm{\Phi} - \widetilde{\bm{X}} \bm{\beta} \|_2^2 + \mathcal{R}(\bm{\beta}),
	\end{align}
	in which $\mathcal{R}(\bm{\beta})$ is a regularization. Since we consider the case $N_k \gg N_o$, $\widetilde{\bm{X}}$ is a rank-deficient matrix and a regularization is required to guarantee the uniqueness of the solution. We refer to (\ref{eqn:optimization-1}) as a least-square (LS) inverse model.
	
\subsection{Adjoint model}\label{sec:adjoint}
	It is important to note that computing $\widetilde{\bm{X}}$ requires to solve the advection-diffusion equation for $N_k$ times, which makes it computationally impractical as $N_k$ becomes large. Moreover, as will be discussed in section \ref{sec:uncertainty}, when the model uncertainty is considered, the total number of computation easily blows up to $O(10^3 \sim 10^4)$. To circumvent these difficulties, an adjoint model is employed in this study. Reducing the number of repetitive computations from the number of GRBFs, $N_k$, to the number of observations, $N_o$, an adjoint model is computationally more tractable \citep{Keats07}.
	
	Here, the adjoint model is briefly described. Define a conjugate field ($\phi^*$) as
	\begin{equation} \label{eqn:dual}
	\begin{cases}
	\langle \phi(\bm{x},t),\chi(\bm{x},t;\bm{x}_i^o,t^o) \rangle = \langle \phi_i^*(\bm{x},t;t_o),Q(\bm{x}) \rangle & \text{for}~i = 1,\cdots,N_o \\
	\phi^*_i(\bm{x},t) = 0 & \text{for}~t \ge t^o
	\end{cases}
	\end{equation}
	Then, the adjoint operator is obtained from the Lagrangian duality relation; 
	\begin{equation}
	\langle \mathcal{A}\phi, \phi_i^* \rangle = \langle \phi, \mathcal{A}^* \phi_i^* \rangle =\langle \phi(\bm{x},t),\chi(\bm{x},t;\bm{x}_i^o,t^o) \rangle,
	\end{equation}
	which gives
	\begin{equation} \label{eqn:governing_adjoint}
	\mathcal{A}^*\phi^*_i(\bm{x},t;t^o) = -\frac{\partial}{\partial t}\phi_i^* - \bm{u}\cdot\nabla \phi_i^* - \nabla \cdot (\bm{K} \cdot \nabla \phi^*_i) = \chi_i,~~\text{for}~t\in(-\infty,t^o),
	\end{equation}
	for $\chi_i = \chi(\bm{x},t;\bm{x}_i^o,t^o)$. The adjoint model (\ref{eqn:governing_adjoint}) is solved backward in time from $t^o$. For more details, see \cite{Pudykiewicz98}.
	
	Once the $i$-th conjugate field $\phi_i^*$ is computed by solving (\ref{eqn:governing_adjoint}) with appropriate boundary conditions \citep{Hourdin06}, $\phi$ at the $i$-th sensor is computed as
	\begin{equation}
	\langle \phi(\bm{x},t),\chi(\bm{x},t;\bm{x}_i^o,t^o) \rangle = \langle \phi_i^*(\bm{x},t;t_o),Q^*(\bm{x}) \rangle = \sum_{j=1}^{N_k} \langle \phi_i^*(\bm{x},t;t_o),\mathcal{P}_j(\bm{x}) \rangle \beta_j.
	\end{equation}
	Repeating the process for $N_o$ conjugate fields, the observation vector is
	\begin{equation} \label{eqn:data-generation}
	\bm{\Phi} = \bm{X} \bm{\beta} + \bm{\epsilon},
	\end{equation}
	in which
	\[
	X_{ij} = \langle \phi_i^*(\bm{x},t;t_o),\mathcal{P}_j(\bm{x}) \rangle.
	\]
	It is trivial to show that $\widetilde{\bm{X}} = \bm{X}$. 
	The coefficients $\bm{\beta}$ can be computed by solving the same least-square minimization problem (\ref{eqn:optimization-1}).
	
		
\section{Generalized polynomial chaos for model uncertainty} \label{sec:uncertainty}
The Gaussian radial basis function, $\bm{\mathcal{P}}$, distributes the source strength $\bm{\beta}$ computed from (\ref{eqn:optimization-1}) in the space centered on the collocation points of $\mathcal{W}$. Obviously, the solution $Q^*(\bm{x})$ of a LS inverse model depends on the choice of $\mathcal{W}$. For example, if a local peak of $Q(\bm{x})$ does not coincide with one of the collocation points, the LS inverse model will result in a poor accuracy. In general, there is no standard rule of choosing $\mathcal{W}$. In this study, we propose to represent the uncertainty in the choice of $\mathcal{W}$ as a random variable. 

Let $\mathcal{W}^*(\omega)$ be a random variable with a uniform distribution; 
	\begin{equation} \label{eq:random_colloc}
	\mathcal{W}^*(\omega) = (\bm{s}_1 + \xi_1(\omega) \Delta) \otimes (\bm{s}_2 + \xi_2(\omega) \Delta).
	\end{equation}
	Here, $\xi_1(\omega)$ and $\xi_2(\omega)$ are real random variables defined over a probability space $(\Omega,\mathcal{S},\mathcal{P})$, in which $\Omega$ is the sample space, $\mathcal{S}$ is the $\sigma$-algebra, $\mathcal{P}$ is the probability measure, and $\omega$ is an element of $\Omega$. Both $\xi_1(\omega)$ and $\xi_2(\omega)$ are defined over $\Gamma_1 = \Gamma_2 = (-0.5,0.5)$ with the probability density functions, $\rho_i(\xi):\Gamma_i \rightarrow \mathbb{R}^+$, $\rho_1(\xi_1) = \rho_2(\xi_2) = 1$, \ie $\xi_i \sim \mathcal{U}(-0.5,0.5)$ for $i = 1$, 2. 
Note that $\mathcal{W}^*(\omega)$ corresponds to a random translation of $\mathcal{W}$, which uniformly covers the entire computational domain.
In the absence of prior information on the source location, a natural choice would be to give an equal probability to every possible $\mathcal{W}$.
	
	On the random collocation system, $\mathcal{W}^*(\omega)$,  (\ref{eqn:data-generation}) becomes
	\begin{equation} \label{eqn:data-uq}
	\bm{\Phi} = \bm{X}(\omega) \bm{\beta}(\omega) + \bm{\epsilon},
	\end{equation}
	in which
	\[
	X_{ij}(\omega) = \langle \phi_i^*(\bm{x},t;t_o),\mathcal{P}_j(\bm{x};\omega) \rangle,
	\]
	and
	\[
	\mathcal{P}_i(\bm{x};\omega) = \frac{1}{2\pi (c \Delta)^2} \exp \left[ - \frac{1}{2} \frac{ |\bm{x}-(\bm{y}^i+\Delta \bm{\xi}(\omega))|^2}{(c \Delta)^2} \right].
	\]
Here, we aim to model $Q(\bm{x})$ by the first moment of the stochastic system. Although it is possible to develop a model matching higher moments, it will lead to a complex non-convex optimization problem.
Taking an expectation over $\bm{\xi}$, (\ref{eqn:data-uq}) becomes
	\begin{equation}
	\bm{\Phi} = E_{\bm{\xi}}[\bm{X}(\omega) \bm{\beta}(\omega)] + \bm{\epsilon}.
	\end{equation}
	Then, a least-square inverse model can be formulated as
	\begin{align} \label{eqn:optimization-2}
	\underset{{\bm{\beta}(\omega)}}{\text{arg min}} \frac{1}{2} \| \bm{\Phi} - E_{\bm{\xi}}[ \bm{X}(\omega) \bm{\beta}(\omega) ] \|_2^2 + \mathcal{R}(\bm{\beta}(\omega)),~~\text{s.t.}~~E_{\bm{\xi}}[Q^*(\bm{x};\omega)] \ge 0~~\forall \bm{x} \in D,
	\end{align}
	in which
	\[
	Q^*(\bm{x};\omega) =  \sum_{i=1}^{N_k} \beta_i(\omega) \mathcal{P}_i(\bm{x};\omega).
	\]
Note that, since we consider only the first moment, the non-negative condition is imposed only on the expectation of $Q^*(\bm{x};\omega)$.
	Hereafter, the obvious dependence on $\bm{\xi}$ is omitted in the expectation, i.e., $E[f] = E_{\bm{\xi}}[f]$. 
	
	Following \cite{Xiu07,Xiu02}, the generalized polynomial chaos (gPC) expansion is employed to approximate the stochastic functions;
	\begin{align}
	\bm{\beta}(\bm{\xi}) &= \sum_{i=0}^{M} \widehat{\bm{\beta}}^i \Psi^i(\bm{\xi}), \label{eqn:gPC-beta}\\
	\bm{X}(\bm{\xi}) &= \sum_{i=0}^{M} \widehat{\bm{X}}^i \Psi^i(\bm{\xi}). \label{eqn:gPC-X}
	\end{align}
	Here, $\Psi^i(\bm{\xi})$ is an orthonormal polynomial basis in a bivariate polynomial space ($Z_2^P$) constructed by a tensor product of one-dimensional polynomial spaces
	\[
	Z_2^P \equiv \bigotimes_{i=1}^2 W^{i,P},
	\]
	in which 
	\[
	W^{i,P} \equiv \left\{ v: \Gamma_i \rightarrow \mathbb{R}: v \in \text{span}\{\psi_m(\xi_i)\}_{m=0}^P\right\}.
	\]
	As $\bm{\xi}$ is a uniform random variable, the Legendre polynomial is chosen as the basis polynomial $\psi_m(\xi)$ \citep{Xiu02}. 
	
	\begin{figure}
		\centering
		\includegraphics[width=0.45\textwidth]{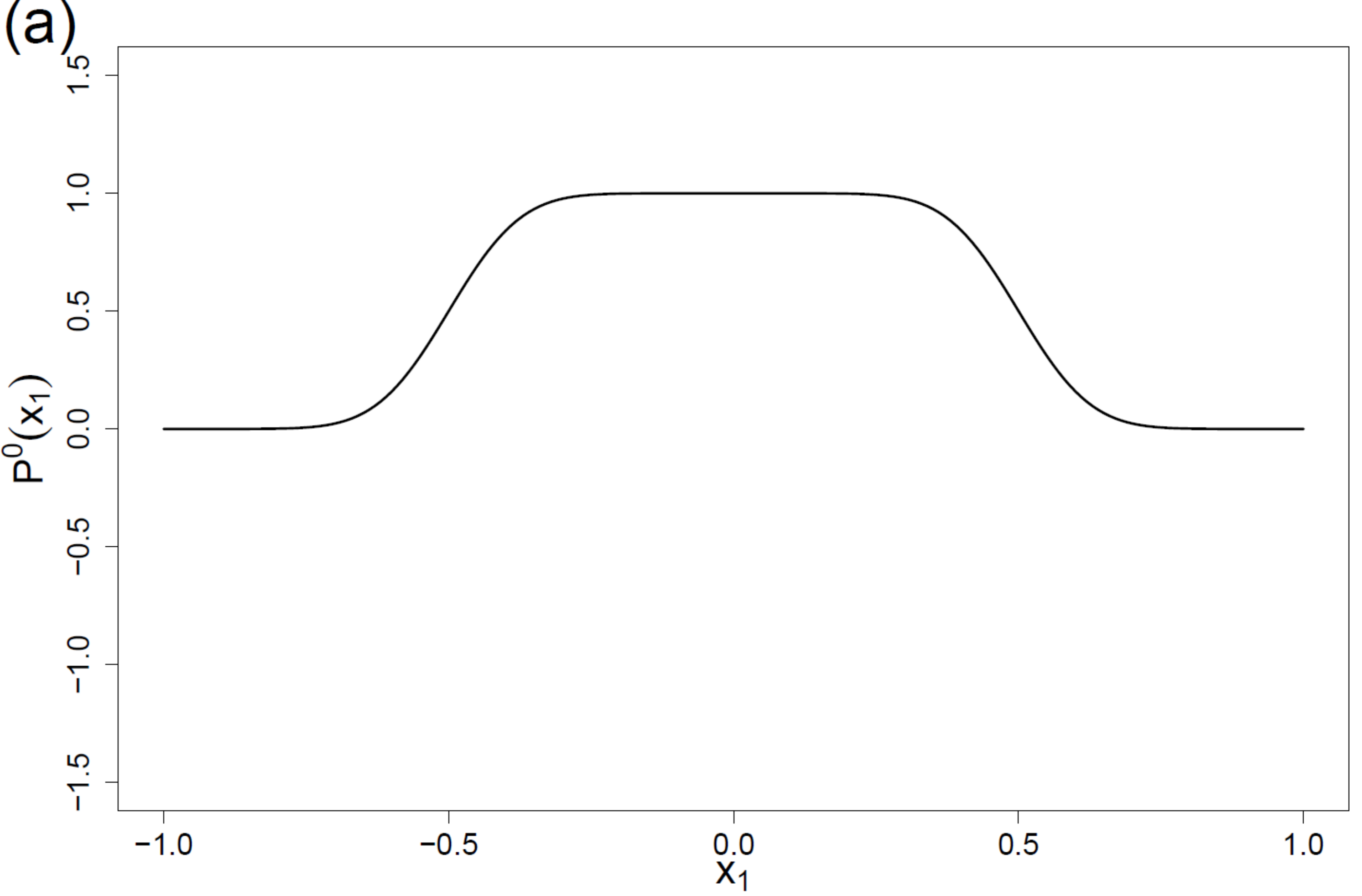}~
		\includegraphics[width=0.45\textwidth]{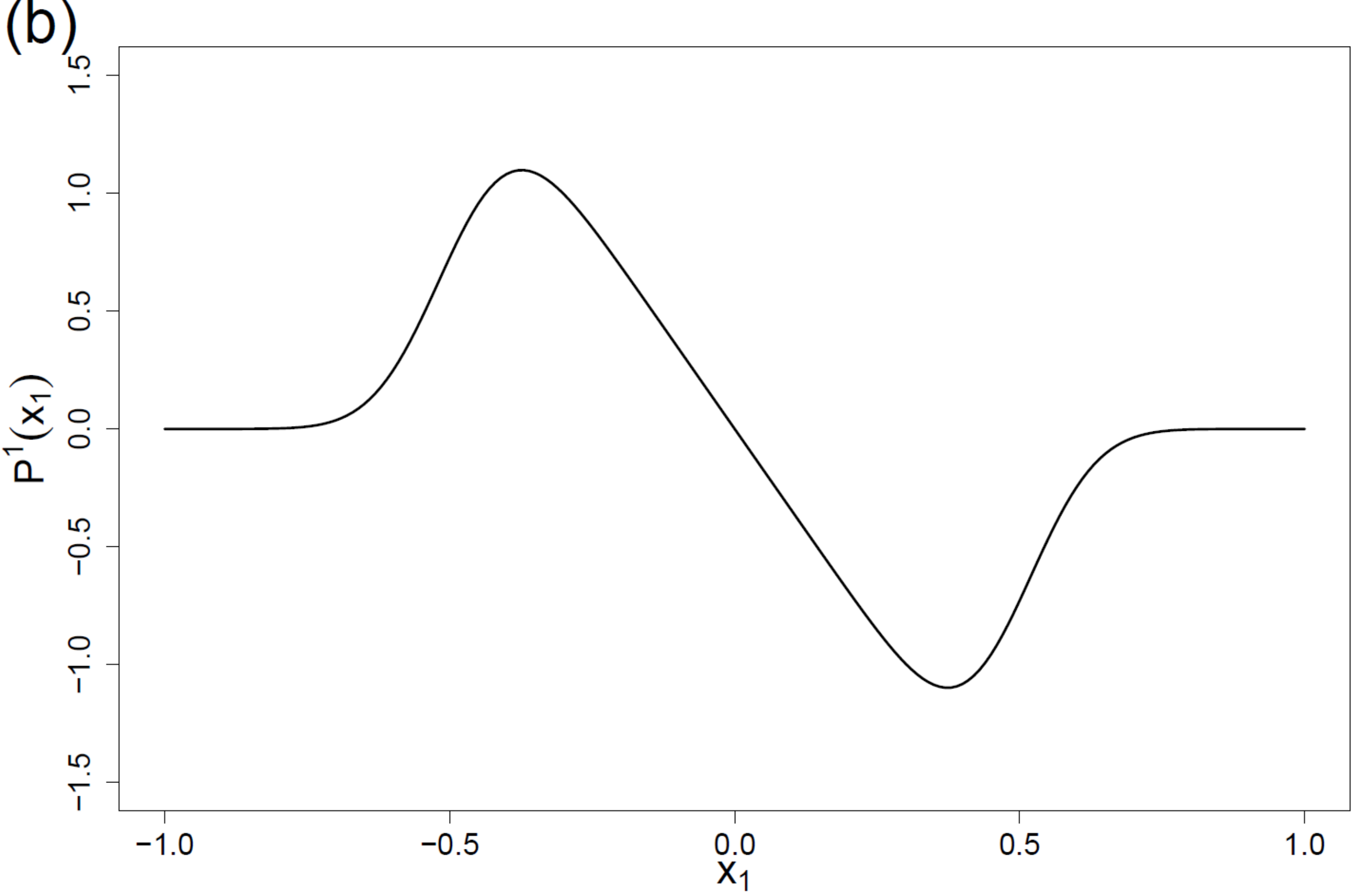}\\
		\includegraphics[width=0.45\textwidth]{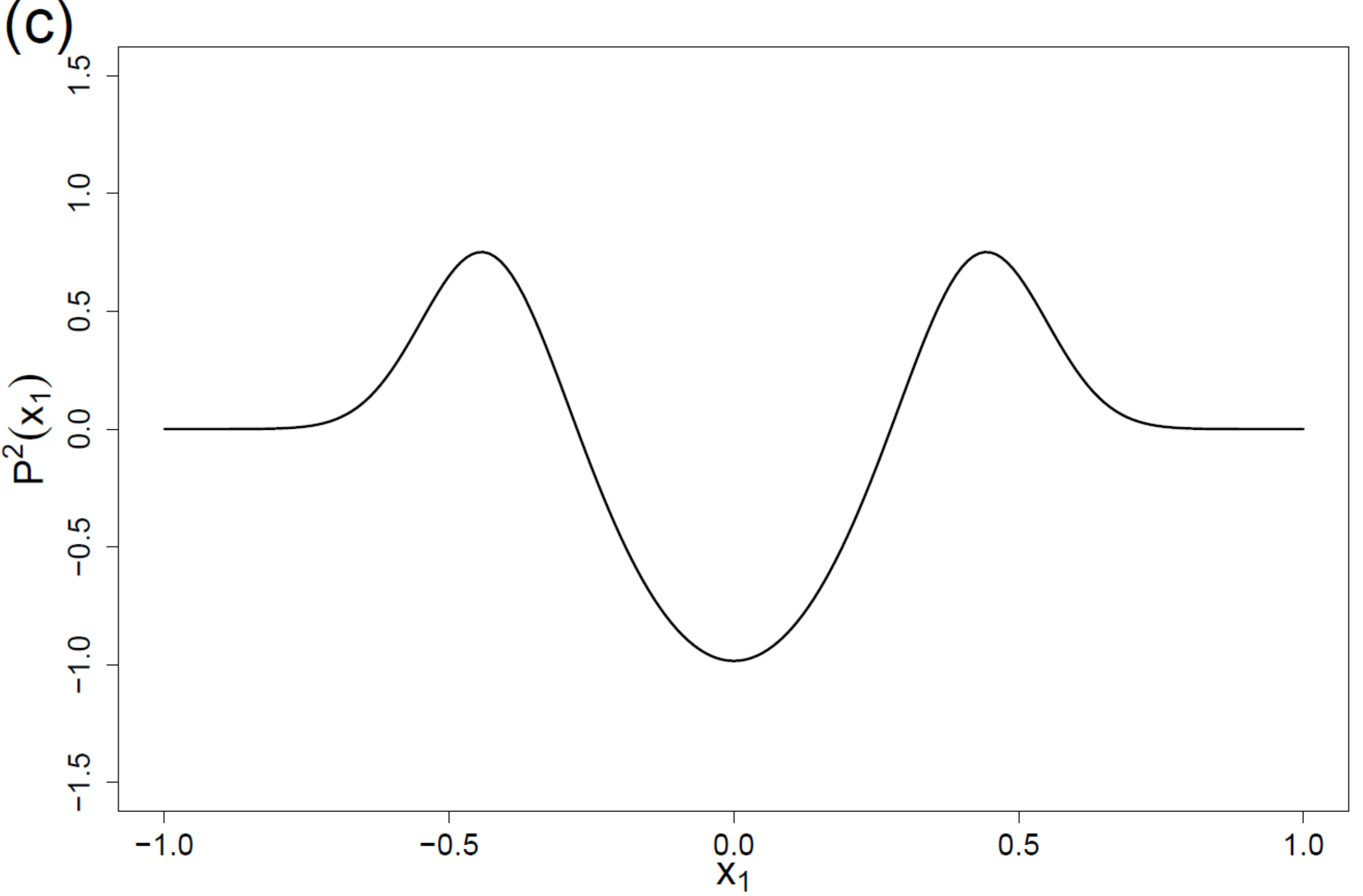}~
		\includegraphics[width=0.45\textwidth]{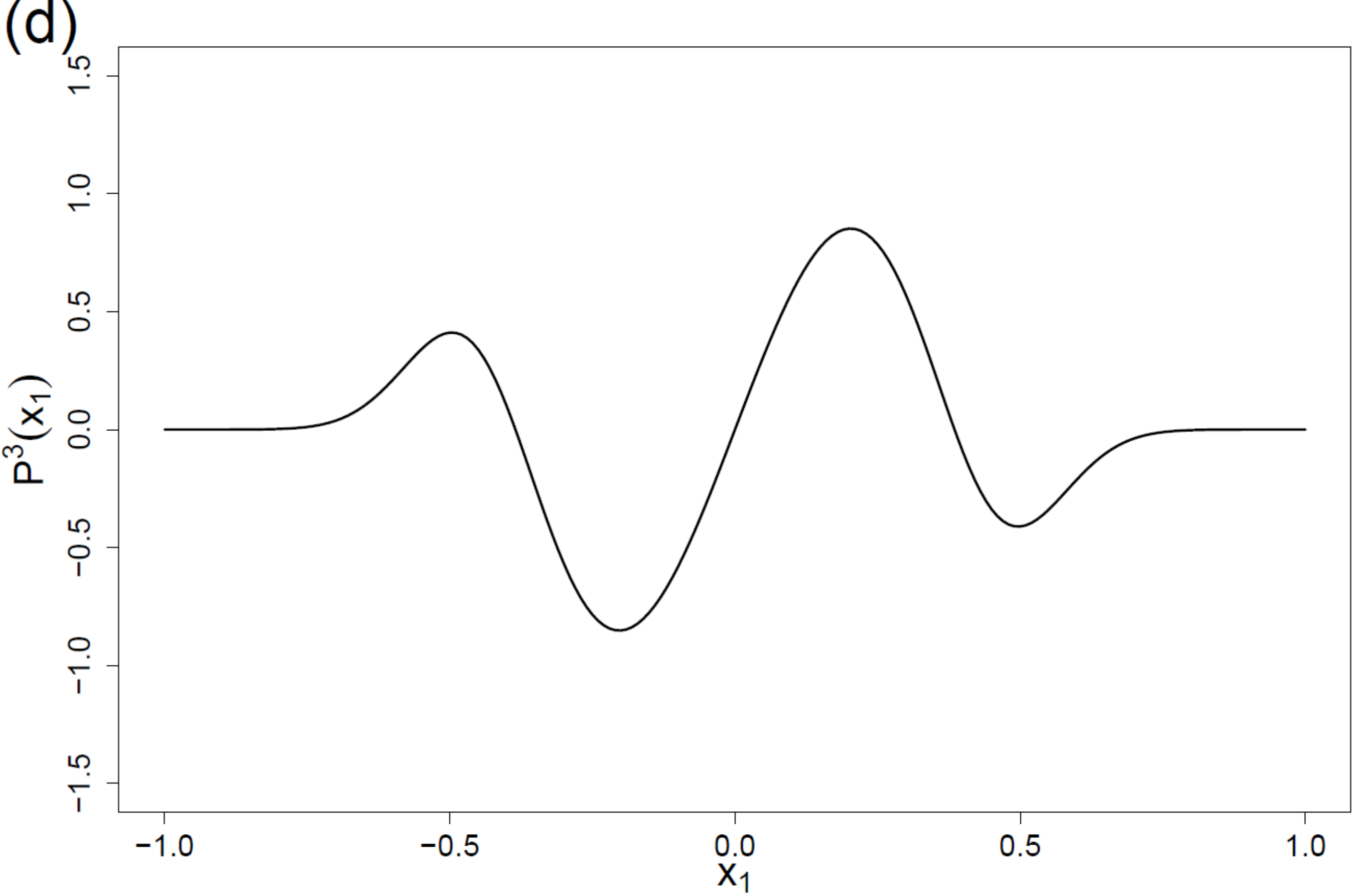}\\
		\caption{ First four modes of the one-dimensional gPC coefficients, $\widehat{\mathcal{P}}^i(x_1)$, for $c = 0.1$. $x_1$ is normalized by $\Delta$. }\label{fig:basis}
	\end{figure}
	
	From the gPC approximation,  
	\begin{equation}
	E[ \bm{X}(\bm{\xi}) \bm{\beta}(\bm{\xi}) ] = \sum_{i=0}^M \sum_{j=0}^M \widehat{\bm{X}}^i \widehat{\bm{\beta}}^j  \int_{\bm{\xi}\in\Gamma} \Psi^i(\bm{\xi})\Psi^j(\bm{\xi})\rho(\bm{\xi}) d\bm{\xi} = \sum_{i=0}^M \widehat{\bm{X}}^i \widehat{\bm{\beta}}^i ,
	\end{equation}
	in which $\Gamma = \Gamma_1 \times \Gamma_2$, $\rho(\bm{\xi})=\rho_1(\xi_1) = \rho_2(\xi_2) = 1$, and $M$ denotes the total number of the basis functions excluding the mean (zero-th order) component, $M = (P+1)^2-1$. 
Here,
	\begin{equation}  \label{eqn:X_hat}
	\widehat{X}^i_{lm} = \langle \phi_l^*(\bm{x},t;t_o),\widehat{\mathcal{P}}^i_m(\bm{x}) \rangle.
	\end{equation}
	And, the source function is
	\begin{equation}
	E[ Q^*(\bm{x};\bm{\xi})] =   \sum_{i=1}^{N_k} \left( \sum_{j=0}^M \widehat{\beta}^j_i \widehat{\mathcal{P}}^j_i(\bm{x}) \right).
	\end{equation}
	The gPC mode of GRBF, $\widehat{\mathcal{P}}^i(\bm{x})$, is
	\begin{equation} \label{eqn:P_hat}
	\widehat{\mathcal{P}}^i_j(\bm{x}) = \frac{1}{2\pi (c \Delta)^2}  \int \exp \left[ - \frac{1}{2} \frac{ |\bm{x}-(\bm{y}^j+\Delta \bm{\xi})|^2}{(c \Delta)^2} \right] \Psi^i(\bm{\xi}) d\bm{\xi}.
	\end{equation}
	The coefficients $\widehat{\bm{\mathcal{P}}}$ can be easily computed by using a numerical integration such as the Gaussian quadrature. 
	For low order modes, $\widehat{\bm{\mathcal{P}}}$ can be even computed analytically. For example, the first mode of the expansion is
	\begin{align} \label{eqn:P0}
	\widehat{\mathcal{P}}^0_j(\bm{x}) &= \frac{1}{2\pi (c \Delta)^2}  \int \exp \left[ - \frac{1}{2} \frac{ |\bm{x}-(\bm{y}^j+\Delta \bm{\xi})|^2}{(c \Delta)^2} \right] \Psi^0(\bm{\xi}) d\bm{\xi}\\
	&= \frac{1}{4\Delta^2} \prod_{i=1}^2 \left[ erf\left( \frac{y^j_i-x_i+0.5\Delta}{\sqrt{2}c} \right) - erf\left( \frac{y^j_i-x_i-0.5\Delta}{\sqrt{2}c} \right) \right].  \nonumber
	\end{align}
	Figure \ref{fig:basis} shows the first four modes of $\widehat{\mathcal{P}}^i(\bm{x})$. It is shown that $\widehat{\bm{\mathcal{P}}}(\bm{x})$ constitutes spatial hierarchical basis functions to approximate $Q(\bm{x})$. 
The advantage of using the hierarchical basis functions over increasing the number of GRBF ($N_k$) is discussed in section \ref{sec:optimization}.
	
\section{Regularized optimization formulation}\label{sec:optimization}
	Using the gPC expansion, the minimization problem (\ref{eqn:optimization-2}) becomes,
	\begin{align} \label{eqn:optimization-3}
	\underset{{\widehat{\bm{\beta}} \in \mathbb{R}^{N_k (M+1)} }}{\text{arg min}} \frac{1}{2} \| \bm{\Phi} - \sum_{i=0}^M \widehat{\bm{X}}^i \widehat{\bm{\beta}}^i  \|_2^2 + \mathcal{R}(\widehat{\bm{\beta}}),~~\text{s.t.}~~\sum_{i=1}^{N_k} \sum_{j=0}^M  \widehat{\beta}^j_i \widehat{\mathcal{P}}^j_i(\bm{x}) \ge 0~~\forall \bm{x} \in D.
	\end{align}
	In the minimization problem, the number of parameters to estimate is $N_{\beta} = N_k \times (M+1)$.  
	Note that, if we use the forward simulation approach shown in section \ref{sec:forward}, even for $N_k = O(100)$ and $P = 10$, the total number of numerical simulations to compute $\widehat{\bm{X}}$ becomes $O(10^4)$. On the other hand, using the adjoint model, the total number of the numerical simulations remains as $N_o$, which is $\sim O(10)$ and $\widehat{\bm{X}}$ can be computed efficiently by evaluating the inner product $\langle \phi^*, \widehat{\mathcal{P}}(\bm{x}) \rangle$ with a numerical integration.
	
	In (\ref{eqn:optimization-3}), $\widehat{\bm{\beta}}$ should satisfy the non-negativity constraint, $E[Q^*(\bm{x})]\ge 0$ for every $\bm{x} \in D$. Because it is difficult to directly impose the non-negativity condition for every $\bm{x}$, we propose an indirect constraint based on the stochastic collocation approximation \citep{Xiu05}. In the stochastic collocation method, the stochastic functions are approximated by the Lagrange polynomials as
	\begin{align}
	\bm{\beta}(\bm{\xi}) &= \sum_{i=1}^{N} \widetilde{\bm{\beta}}^i L^i(\bm{\xi}), \label{eqn:SC-beta}\\
	\bm{\mathcal{P}}(\bm{x};\bm{\xi}) &= \sum_{i=1}^{N} \widetilde{\bm{\mathcal{P}}}^i(\bm{x}) L^i(\bm{\xi}). \label{eqn:SC-X}
	\end{align}
	Here, $L^i(\bm{\xi})$ is the $i$-th Lagrange polynomial, which is $L^i(\bm{\xi}_j) = \delta_{ij}$ for the $j-$th collocation points in $\Gamma$, $\bm{\xi}_j$. Again, $L(\bm{\xi})$ is constructed by the tensor product of two one-dimensional Lagrange polynomials in $\Gamma_1$ and $\Gamma_2$. The stochastic collocation method corresponds to a deterministic sampling and the coefficients are easily computed by the function evaluations at each collocation points in $\Gamma$, i.e., $\widetilde{\bm{\beta}}^i = \bm{\beta}(\bm{\xi}_i)$ and $\widetilde{\bm{\mathcal{P}}}^i(\bm{x}) = \bm{\mathcal{P}}(\bm{x};\bm{\xi}_i)$. Since GRBF is positivie, $\bm{\mathcal{P}}(\bm{x};\bm{\xi}) \ge 0$ for all $(\bm{x},\bm{\xi}) \in D \times \Gamma$, the non-negativity condition implies $\bm{\beta}(\bm{\xi}_i) \ge 0$ for $i = 1, \cdots, N$. In other words, in the stochastic collocation approach, it is sufficient to impose the non-negativity constraint only on the parameters, not on the field. The stochastic collocation coefficients, $\bm{\beta(\xi_i)}$, are related to the modal gPC coefficients, $\widehat{\bm{\beta}}$, as
	\[
	\bm{\beta}(\bm{\xi}_i) \simeq \sum_{j=0}^M \widehat{\bm{\beta}}^j \Psi^j(\bm{\xi}_i).
	\]
	Then, the constraint on the smooth function surface $E[\bm{Q}^*(\bm{x})] \ge 0$ for every $\bm{x} \in D$ can be approximated by the following linear constraint
	\begin{equation} \label{eqn:beta_constraint}
	\beta_i(\bm{\xi}_j) = \sum_{k=0}^M \widehat{\beta}^k_i \Psi^k(\bm{\xi}_j) \ge 0~~~\text{for}~i=1,\cdots,N_k~\&~j=1,\cdots,N.
	\end{equation}
	Or,
	\begin{equation}
	\mathcal{L}\widehat{\bm{\beta}} \ge 0,
	\end{equation}
	in which $\mathcal{L}$ is a block diagonal matrix for $\Psi^i(\bm{\xi}_j)$. In this study, we choose $N = [\frac{3}{2}(P+1)]^2$ and (\ref{eqn:beta_constraint}) is evaluated at the Chebyshev node.
		
	 As an analogy to the finite element analysis \citep{Karniadakis05}, either $h$- or $p$-type refinement can be used to increase the resolution of the proposed inverse model. Let $N_k$ be the number of GRBFs for a reference case, i.e., the number of collocation points of $\mathcal{W}$. In the $h$-type refinement, the grid space is decreased as $\Delta_q = \Delta /q$ for $q\in\mathbb{N}^+$, which makes the total number of unknown parameters $N_{\beta} = N_k \times q^2$. In the $p$-type refinement, $N_k$ is fixed and the maximum order of $\Psi(\bm{\xi})$ is increased, which results in $N_{\beta}=N_k \times p^2$ for $p = P+1$. Because of this quadratic dependence,  in both $h$- and $p$-type refinements, the number of unknown parameters can easily overwhelm the number of observations upon a refinement. 
For example, in the numerical experiments in section \ref{sec:results}, the number of unknown parameters, i.e. the dimension of $\widehat{\bm{\beta}}$, is $\sim O(10^3-10^4)$, while the number of data $\sim O(10)$. 
As a result, the optimization formulation results in a highly ill-posed system, of which solution heavily relies on the choice of the regularization.
To alleviate the difficulty, we develop a regularization strategy, which exploits the hierarchical nature of the GRBF-gPC coefficients, $\widehat{\bm{\mathcal{P}}}(\bm{x})$, 
	 
As shown in (\ref{eqn:P0}), the zero-th gPC mode of GRBF, $\widehat{\bm{\mathcal{P}}}^0_j(\bm{x})$, represents the average source strength in $(-0.5\Delta,0.5\Delta)\times(-0.5\Delta,0.5\Delta)$ centered at $\bm{y}^i$. Because $\widehat{\bm{\mathcal{P}}}^0_j(\bm{x}) \ge 0$, the non-negativity constraint implies that 
	\begin{equation}
	\widehat{\beta}^0_i \ge 0 ~\text{for}~i = 1,\cdots,N_k.
	\end{equation}
At the same time, when the mean source strength, $\widehat{\beta}^0_i$, is zero, the variation around the mean, represented by the higher-order gPC modes, should also be zero, i.e., 	
	\begin{equation} \label{eqn:non-negative_const}
	\widehat{\beta}^k_i = 0~\text{for every}~k \ge 1,~\text{if}~ \widehat{\beta}^0_i = 0.
	\end{equation}
Therefore, in the $p$-type refinement, the number of unknown parameters can be effectively reduced by identifying non-zero elements in $\widehat{\bm{\beta}}^0$. 

	From these observations, we propose the following mixed $l_1$- and $l_2$-regularizations,
	\begin{equation}
	\mathcal{R}(\widehat{\bm{\beta}}) = \lambda_1 \| \widehat{\bm{\beta}}^0 \|_1 + \lambda_2 \| \widehat{\bm{\beta}}' \|^2_2,
	\end{equation} 
	in which $\widehat{\bm{\beta}}' = ( \{\widehat{\bm{\beta}}^1\}^T,\cdots,\{\widehat{\bm{\beta}}^M\}^T )^T$. The Least Absolute Shrinkage and Selection Operator (LASSO), or $l_1$-regularization, is one of the most widely used regularization methods to find such a ``sparse'' solution for the so-called ``large $m$, small $n$'' problem (large number of parameters and small number of data) \citep{Tibshirani1996}. As discussed above, for the $p$-type refinement, applying LASSO only for $\widehat{\bm{\beta}}^0$ is enough to guarantee a sparse solution. Hence, LASSO is applied only to the zeroth mode, $\widehat{\bm{\beta}}^0$, and the higher-order terms are regularized by the standard Tikhonov regularization. 
	
	From the non-negativity constraint, we know that $\widehat{\bm{\beta}}^0 \in \mathbb{R}_{\ge 0}^{N_k}$. Then, the regularized optimization problem can be written as
	\begin{align} \label{eqn:optimization-4}
	&\underset{\bm{\beta}(\bm{\xi})}{\text{arg min}} \frac{1}{2} \| \bm{\Phi} - E[\bm{X(\bm{\xi}) \beta(\bm{\xi})}]  \|_2^2 +  \lambda_1 E[\bm{\beta}(\bm{\xi})] + \lambda_2 tr\left( Cov\left[\bm{\beta}(\bm{\xi}),\bm{\beta}(\bm{\xi}) \right] \right), \\
	&\text{s.t.}~~E[Q^*(\bm{x})] \ge 0,~~\forall \bm{x} \in D. \nonumber
	\end{align}
	The first tuning parameter $\lambda_1$ controls the sparsity in the solution, $\bm{\beta}$, and the second one $\lambda_2$ prevents the overfitting by regularizing the total variation around the mean.
	
	Furthermore, to consider the spatial smoothness of $Q(\bm{x})$, the fused LASSO is used \citep{Tibshirani2005}. In the fused LASSO, $l_1$-norm is imposed on the difference between a directly connected parameters. For example, in the $x_1$-direction, the fused LASSO regularization is
	\begin{equation}
	\|\bm{G}_1\widehat{\bm{\beta}}^0\|_1 = \sum_{(i,j) \in \mathcal{N}_1} | \widehat{\beta}^0_i - \widehat{\beta}^0_j |,
	\end{equation}
	in which $\mathcal{N}_1$ is an index set for $\widehat{\bm{\beta}}^0$ directly connected in the $x_1$-direction, i.e.,
	\[
	\mathcal{N}_1 = \{(i,j): (\bm{y}^i - \bm{y}^j)\cdot \bm{e}_1 = \Delta~\text{for}~i = 1,\cdots,N_k-1~\text{and}~ i < j \le N_k\}.
	\] 
	In other words, the fused LASSO is equivalent to imposing $l_1$-penalty in the gradient of $Q^*(\bm{x})$. We can define the difference matrix in the $x_2$-direction, $\bm{G}_2$, similar to $\bm{G}_1$. Then, the regularization can be written as a mixed generalized LASSO \citep{Tibshirani2011} and Tikhonov regularization,
	\begin{equation}
	\mathcal{R}(\widehat{\bm{\beta}}) = \lambda_1 \| \bm{S} \widehat{\bm{\beta}}^0 \|_1 + \lambda_2 \| \widehat{\bm{\beta}}' \|^2_2,
	\end{equation} 
	in which
	\[
	\bm{S} = 
	\begin{bmatrix}
	\gamma \bm{I} \\ \bm{G}_1 \\ \bm{G}_2
	\end{bmatrix}.
	\]
	The coefficient $\gamma$ determines the relative weight of the standard LASSO to the fused LASSO.
	
	Finally, the optimization problem for the $hp$-type source reconstruction is
	\begin{align} \label{eqn:optimization-final}
	\underset{{\widehat{\bm{\beta}} \in \mathbb{R}^{N_k (M+1)} }}{\text{arg min}} \frac{1}{2} \| \bm{\Phi} - \sum_{i=0}^M \widehat{\bm{X}}^i \widehat{\bm{\beta}}^i  \|_2^2 + \lambda_1 \| \bm{S} \widehat{\bm{\beta}}^0 \|_1 + \lambda_2 \| \widehat{\bm{\beta}}' \|^2_2,~~\text{s.t.}~~\mathcal{L}\widehat{\bm{\beta}} \ge 0.
	\end{align}
Note that the coupling between the modes shown in (\ref{eqn:non-negative_const}) is not explicitly imposed in the regularization. However, the sparsity in the modal domain is implicitly imposed through the linear constraint. This optimization problem is solved by using the Alternating Direction Method of Multipliers (ADMM). ADMM is one of the most widely used method for a large-scale optimization \citep{ADMM2010}. 
	
	Rewriting (\ref{eqn:optimization-final}) as a constraint optimization problem,
	\begin{align} 
	\underset{{\widehat{\bm{\beta}} \in \mathbb{R}^{N_k (M+1)} }}{\text{arg min}} \frac{1}{2} \| \bm{\Phi} - \mathcal{\bm{X}} \widehat{\bm{\beta}}  \|_2^2 + \lambda_1 \| \bm{\alpha} \|_1 + \lambda_2 \| \widehat{\bm{\beta}}' \|^2_2,~~\text{s.t.}~~\bm{F}\widehat{\bm{\beta}} =\bm{\zeta},
	\end{align}
	where  $\mathcal{\bm{X}} = [ \widehat{\bm{X}}^0, \cdots, \widehat{\bm{X}}^M]$, $ \bm{F} = [ \{\bm{S}, \bm{0}\}^\top, \bm{\mathcal{L}}^{\top} ]^{\top}$,   $\bm{\zeta} =(\bm{\alpha}^{\top} , \bm{\theta}_{}^{\top} )^{\top}$, and $\bm{\theta} = (\bm{\mathcal{L}} \widehat{\bm{\beta}})_+  $. Note that $\bm{F}$ is padded with a null matrix $\{\bm{0}\}$, because the generalized LASSO, $\bm{S}$, is applied only to $\widehat{\bm{\beta}}^0$. 
	The subscript $+$ indicates a projection onto a positive set $\mathbb{R}_{\ge0}$, i.e., $(\bm{b})_+ = max(\bm{b},\bm{0})$.
	The augmented Lagrangian form is 
	\begin{align}
	L(\widehat{\bm{\beta}}, \bm{\alpha}, \bm{v}) &= \frac{1}{2} \|  \bm{\Phi} -  \bm{\mathcal{X}} \widehat{\bm{\beta}} \|_2^2 + \lambda_1 \| \bm{\alpha} \|_1+ \lambda_2 \| \widehat{\bm{\beta}}' \|^2_2 \\
	&+ \omega \bm{v}^{\top}( {\bm{F}} \widehat{\bm{\beta}} - \bm{\zeta}  )+ \frac{\omega}{2} \| \bm{F}\widehat{\bm{\beta}} - \bm{\zeta} \|_2^2, \nonumber
	\end{align}
	in which $\omega$ is a constant. Then, the solution procedure for the minimization problem is as follows;
	\begin{enumerate}
		\item Set initial conditions for $\widehat{\bm{\beta}}^{(0)}$ and other variables.
		\begin{equation*}
		\widehat{\bm{\beta}}^{(0)} 
		= \{ ( \bm{\mathcal{X}}^{\top} \bm{\mathcal{X}}  + \omega \bm{F}^{\top} \bm{F}  + \lambda_2 \widetilde{\bm{I}})^{-1}( \bm{\mathcal{X}}^{\top}\bm{\Phi} ) \}_+.
		\end{equation*}
		Here, $\widetilde{\bm{I}}$ is an identity matrix of the dimension of $\widehat{\bm{\beta}}$, whose first $N_k$ elements are zero, which imposes the Tikhonov regularization on $\widehat{\bm{\beta}}'$. Then,
		\begin{eqnarray}
		\bm{\alpha}^{(0)} &=& \bm{S}\widehat{\bm{\beta}}^{0,\,(0)}, \nonumber \\
		\bm{\theta}^{(0)} &=& (\bm{\mathcal{L}}\widehat{\bm{\beta}}^{(0)})_+, \nonumber \\
		\bm{v}^{(0)} &=& \bm{0}. \nonumber
		\end{eqnarray}
		
		\item For $t > 0$, update $\widehat{\bm{\beta}}^{(t+1)}$ by
		\begin{equation*}
		\widehat{\bm{\beta}}^{(t+1)} 
		= ( \bm{\mathcal{X}}^{\top} \bm{\mathcal{X}}  + \omega \bm{F}^{\top} \bm{F}  + \lambda_2 \widetilde{\bm{I}})^{-1}( \bm{\mathcal{X}}^{\top}\bm{\Phi} + \omega  \bm{F}^{\top} ( \bm{\zeta}^{(t)} - \bm{v}^{(t)} ) ).
		\end{equation*}
		
		\item Next, update $\bm{\zeta}$, i.e., $\bm{\alpha}$ and $\bm{\theta}$ by 
		\begin{eqnarray}
		\bm{\alpha}^{(t+1)} &=& \text{sign}( [\bm{F}\widehat{\bm{\beta}}^{(t+1)} + \bm{v}^{(t)}]_U)\times(1- (\lambda_1/\omega) / \| [\bm{F}\widehat{\bm{\beta}}^{(t+1)} + \bm{v}^{(t)}]_U \|_2)_+,  \nonumber \\
		\bm{\theta}^{(t+1)} &=& ( [\bm{F}\widehat{\bm{\beta}}^{(t+1)} + \bm{v}^{(t)}]_L )_+, \nonumber \\
		\bm{\zeta}^{(t+1)} &=& 
		\begin{bmatrix}
		\bm{\alpha}^{(t+1)} \\ \bm{\theta}^{(t+1)} \nonumber
		\end{bmatrix}.
		\end{eqnarray}
		Here, $\bm{v}$ is a vector of dimension $N_k + N$, i.e., the number of GRBFs and the number of the quadrature points for the evaluation of the non-negativity condition. The operator $[ \cdot ]_U$ takes only the upper $N_k$ elements of the vector for the generalized LASSO and $[ \cdot ]_L$ subsets the lower $N$ elements of the vector for the non-negativity constraint.
		
		\item Finally, update $\bm{v}$,
		\[
		\bm{v}^{(t+1)} = \bm{v}^{(t)} + (\bm{F} \widehat{\bm{\beta}}^{(t+1)} - \bm{\zeta}^{(t+1)}).
		\]
		
		\item Repeat steps 2 $\sim$ 4, until the improvement
		\[\| \widehat{\bm{\beta}}^{(t+1)} - \widehat{\bm{\beta}}^{(t)}\|_2 < \epsilon_0\]
		for a pre-specified tolerance level $\epsilon_0$.
	\end{enumerate}
The Tikhonov regularization is imposed in Step 2. In Step 3, the fused LASSO is imposed on $\bm{\alpha}$ and the non-negativity constraint is imposed by a projection on the non-negative set in $\bm{\theta}$.
	
\section{Numerical experiments} \label{sec:results}
	
	\begin{figure}
		\centering
		\includegraphics[height=0.4\textwidth]{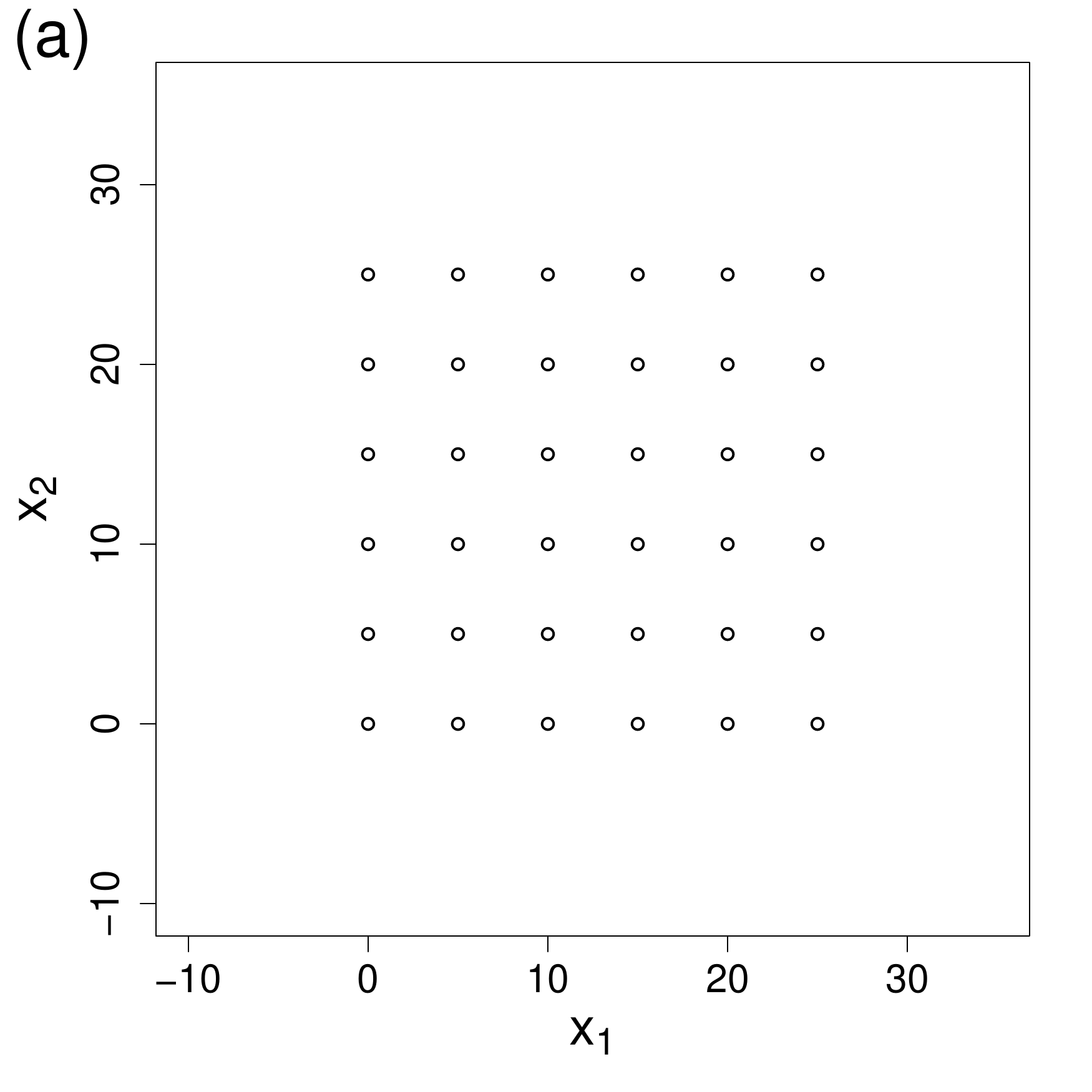}~
		\includegraphics[height=0.4\textwidth]{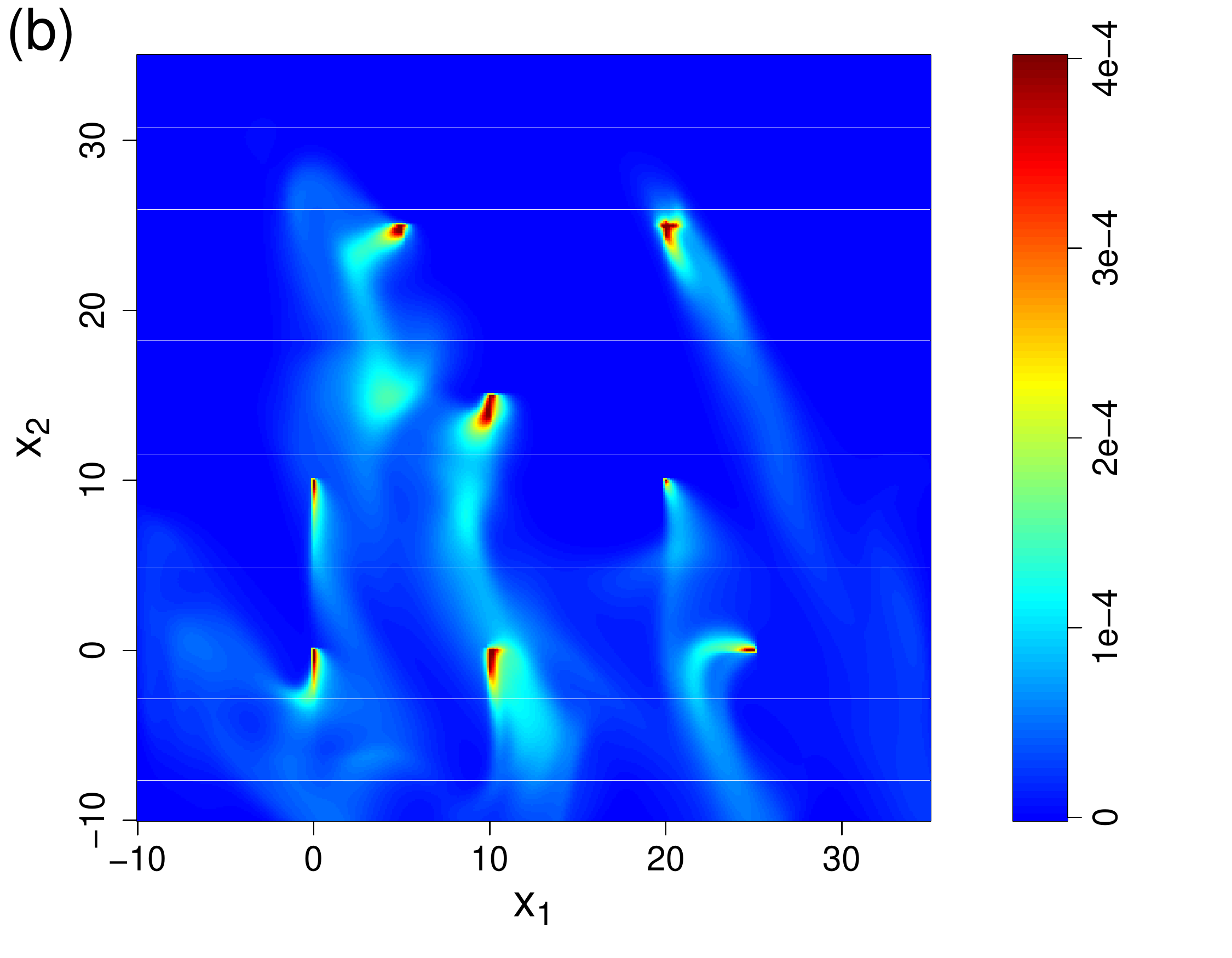}\\
		\caption{ (a) The computational domain and the location of the sensors. (b) Examples of a few time-integrated adjoint field, $\int \phi^*_i(\bm{x},t;t_0)dt$. }\label{fig:domain}
	\end{figure}
	
	We use numerical simulations to study the behavior of the $hp$-inverse model and the results are compared with the standard least-square inverse models (\ref{eqn:optimization-1}) with LASSO and fused-LASSO regularizations \citep{Tibshirani2005,Tibshirani1996}. The computational domain for this case study is $D=(-10,35)\times(-10,35)$. The adjoint equation (\ref{eqn:governing_adjoint}) is numerically integrated for $T=(-5,0)$ by using a third-order low-storage Runge-Kutta method and an upwind finite volume method. The computational grid and time step sizes are chosen as $\delta x_1 = \delta x_2 = 0.2$ and $\delta t =  1/600$. 
	
	Figure \ref{fig:domain} (a) shows the computational domain $D$ as well as the locations of the sensors. The total number of observations is $N_o =  36$. The sensors are located at a rectangular mesh of which grid size is $\Delta_\chi = 5$. The time averaging window of the sensor in equation (\ref{eqn:sensor}) is $T_\chi = 1$ and the time of observation is $t_0 = 0$. The computational parameters are chosen similar to real operational air pollution measurement conditions. In an operational air pollution measurement, usually a high frequency sensor measurement is averaged over one to three hours to remove measurement noise. 
	
	\begin{figure}
		\centering
		\includegraphics[height=0.4\textwidth]{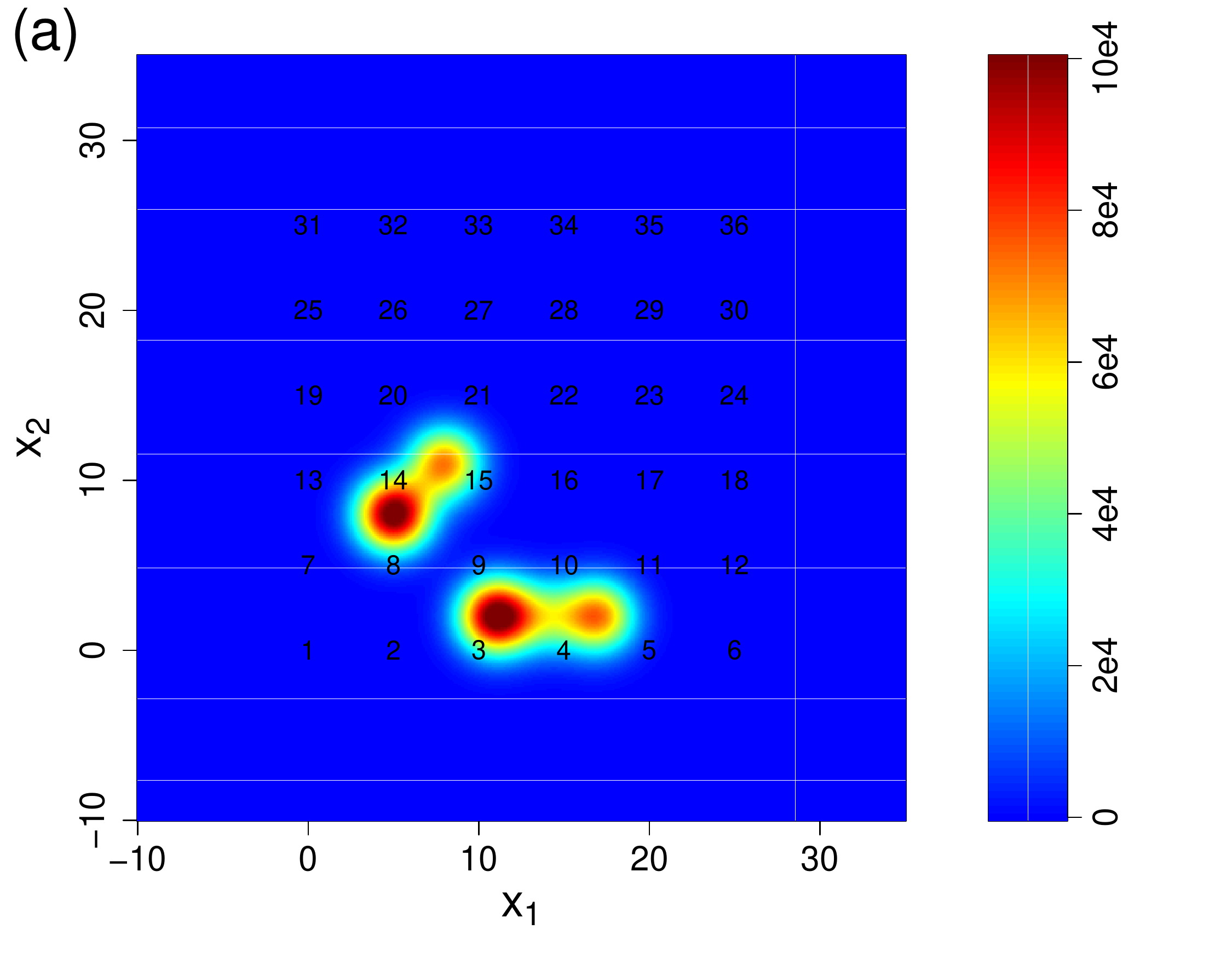}~
		\includegraphics[height=0.4\textwidth]{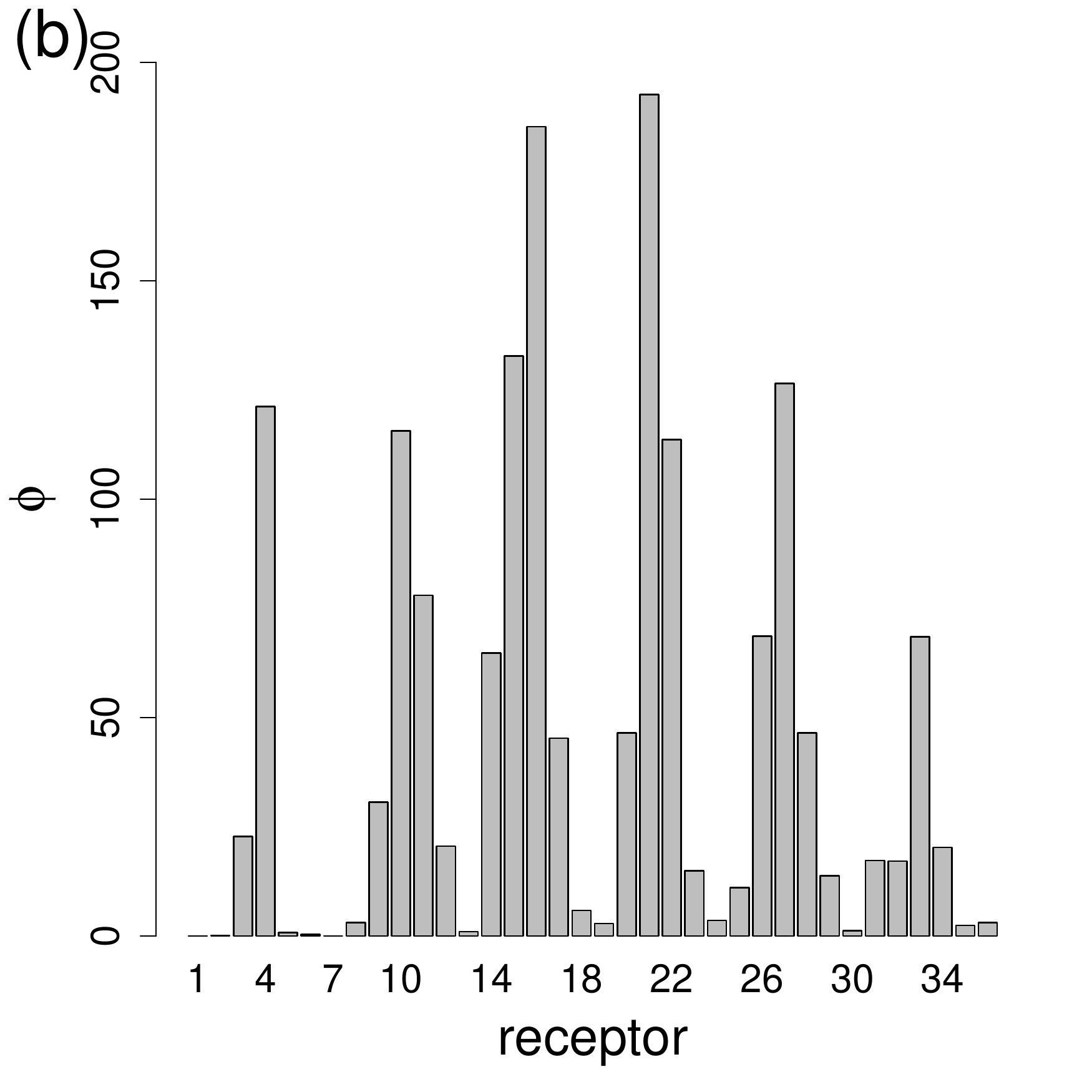}\\
		\caption{ (a) True source surface together with the sensor numbers. (b) The observation ($\bm{\Phi}$) at each sensors. }\label{fig:source_obs}
	\end{figure}
	
	A few adjoint functions are shown in figure \ref{fig:domain} (b). To mimic the atmospheric dispersion process, the wind field is generated by a Fourier transform of a set of Ornstein-Uhlenbeck processes. The velocity in the $x_1$-direction is
	\begin{equation}
	u_1(\bm{x},t) = U^0_1(t) + Real \left(\sum_{l=1}^h \sum_{m=1}^h  (\widehat{u}_r(l,m,t) + i \widehat{u}_i(l,m,t)) e^{i \bm{k}(l,m)\cdot\bm{x}}   \right),
	\end{equation}
	in which $h$ is the maximum number of the Fourier modes and $\bm{k}(l,m) = (\frac{2\pi}{L_1}l,\frac{2\pi}{L_2}m)$ is the wavenumber. The Fourier coefficients are computed by solving the following Langevin equation;
	\begin{equation}
	\delta \widehat{u}_{r,i}(l,m,t) = -\frac{\widehat{u}_{r,i}(l,m,t)}{T_L} \delta t + S(l,m) \delta W,
	\end{equation}
	in which $T_L$ is a relaxation timescale, S is a scale parameter, and $\delta W$ denotes the Wiener process $\sim \mathcal{N}(0,\delta t)$. In this study, $T_L = 2$ and $S=2\sqrt{\frac{2}{T_L (l^2+m^2)}}$ are used \citep{Pope00}. The velocity in the $x_2$-direction is computed from the divergence-free condition;
	\begin{equation}
	\frac{\partial u_1(\bm{x},t)}{\partial x_1} + \frac{\partial u_2(\bm{x},t)}{\partial x_2} = 0.
	\end{equation}
	In other words,
	\begin{equation}
	u_2(\bm{x},t) = U^0_2(t) - Real \left(\sum_{l=1}^p \sum_{m=1}^p  \frac{L_2}{L_1}\frac{l}{m}(\widehat{u}_r(l,m,t) + i \widehat{u}_i(l,m,t)) e^{i \bm{k}(l,m)\cdot\bm{x}}   \right).
	\end{equation}
	The mean components $U^0_{1,2}(t)$ are also obtained by solving the same Langevin equation, but with $S=5\sqrt{\frac{2}{T_L}}$. The diffusivity is computed by an isotropic Smagorinsky model, which is typically used in the atmospheric dynamics models \citep{Byun06}:
	\begin{align}
	&K_{ij}(\bm{x},t) = K_h(\bm{x},t)\delta_{ij}, \nonumber \\
	&K_h(\bm{x},t) = (C_s\Delta_s)^2\sqrt{ \left( \frac{\partial u_1(\bm{x},t)}{\partial x_1} - \frac{\partial u_2(\bm{x},t)}{\partial x_2} \right) +  \left( \frac{\partial u_1(\bm{x},t)}{\partial x_2} + \frac{\partial u_2(\bm{x},t)}{\partial x_1} \right)  }. \nonumber
	\end{align}
	Here, $\delta_{ij}$ is the Kronecker delta, $C_s (= 0.1)$ is the Smagorinsky coefficient, and the length scale $\Delta_s = {\text{max}(L_1,L_2)}/{2\pi h}$. 
The velocity field is generated by one realization of the Ornstein-Uhlenbeck process and assumed to be known. Note that the velocity field is required only when computing $\phi^*$.
	
\subsection{Case study 1}

\subsubsection{Effects of $hp$ refinement}

	Figure \ref{fig:source_obs} (a) shows the true source surface, $Q(\bm{x})$. The data is a snapshot of the concentration observation from the sensor network, shown in figure \ref{fig:source_obs} (b). The goal is to estimate the source surface $Q(\bm{x})$ from the snapshot observation, of which size is $N_o = 36$.


	\begin{figure}
		\centering
		\includegraphics[height=0.32\textwidth]{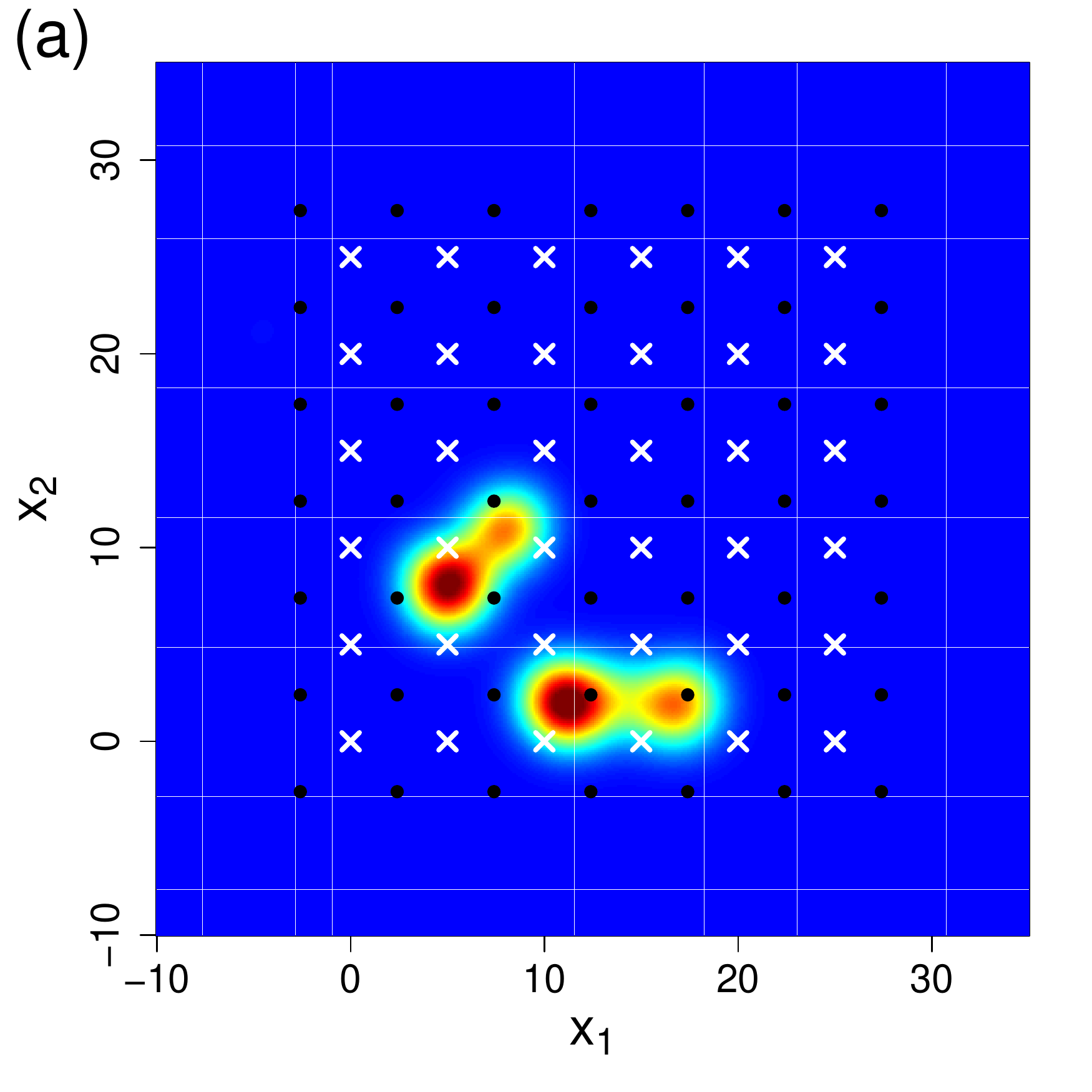}~
		\includegraphics[height=0.32\textwidth]{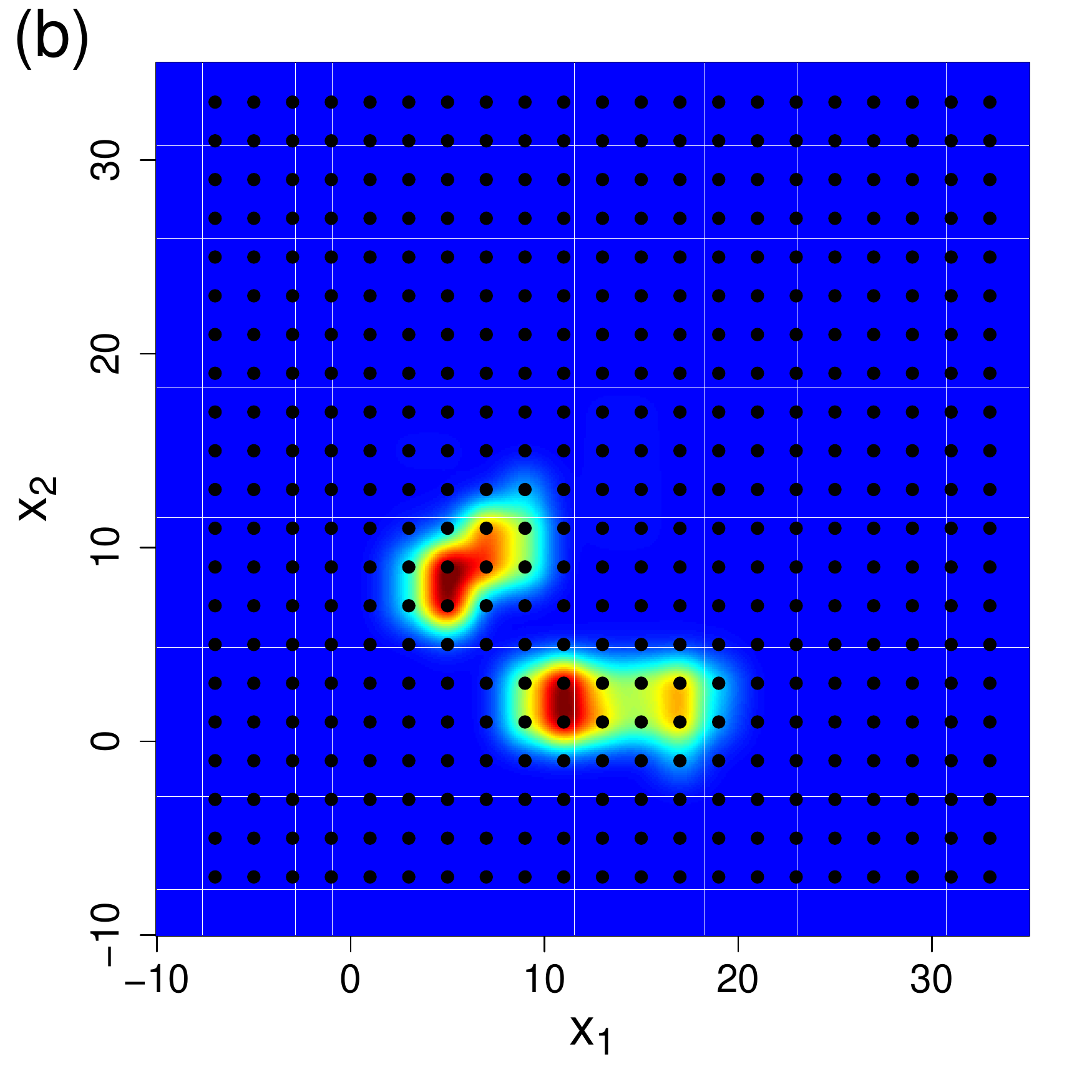}~
		\includegraphics[height=0.32\textwidth]{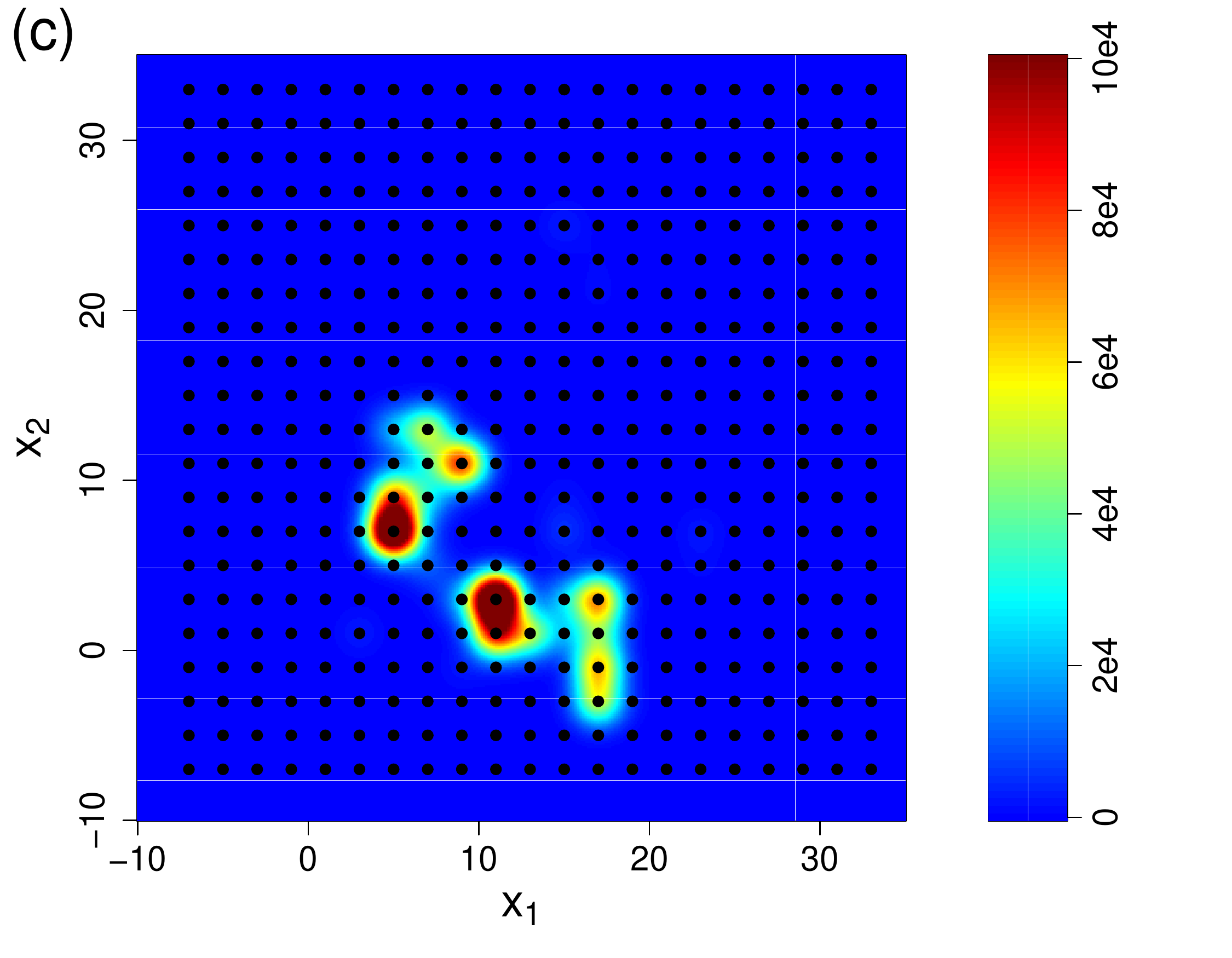}
		\caption{ Source surface estimated by (a) gPC-LASSO ($\Delta = 5$), (b) F-LASSO ($\Delta = 2$), and (c) LASSO ($\Delta = 2$). The black dots indicate the centers of GRBF on $\mathcal{W}$ and the white crosses in (a) are the locations of the sensors.  }\label{fig:comparison_LASSO}
	\end{figure}
	
	Figure \ref{fig:comparison_LASSO} shows the estimated source surface, $Q^*(\bm{x})$, by the proposed $hp$-GRBF inverse model (gPC-LASSO) and the LS inverse models with fused LASSO, and LASSO regularizations. Hereafter, we use F-LASSO and LASSO to refer the LS inverse models with fused LASSO and LASSO regularizations, respectively. In  gPC-LASSO, the grid space of the collocation set $\mathcal{W}$ is $\Delta = 5$ and the scale parameter of GRBF, $\mathcal{P}(\bm{x})$, is set to $c = 0.25$. The maximum order of the Legendre polynomial is $P=5$ in each direction. For F-LASSO and LASSO, $\Delta = 2$ and $c=0.5$ are used. The penalty parameter for $l_1$ regularization is set to $\lambda_1 = 10^{-2}$ for the all three models. For gPC-LASSO, the penalty parameter for $l_2$ regularization is $\lambda_2 = 10^{-6}$. In gPC-LASSO and F-LASSO, $\gamma = 0.5$ is used.
It is  shown that gPC-LASSO provides a better approximation to the true source surface even with a lower resolution GRBF.

\begin{table}
\center{
\caption{ Normalized $l_2$ error.} 
\label{tbl:l2_case1-1}
\begin{tabular}{r|cccc}
\hline \hline
$\Delta$ & 5 & 3 & 2 & 1\\
\hline
gPC-LASSO ($P=5$)& 0.034 & - & - & -\\
F-LASSO & 0.563 & 0.532 & 0.167 & 0.288\\
LASSO & 0.576 & 0.505 & 0.449 & 1.713\\
\hline \hline
\end{tabular}
}
\end{table}

For a quantitative comparison, we define a normalized $l_2$ error,
	\begin{equation}\label{eqn:error}
	e_Q = \frac{\int \{ Q^*(\bm{x}) - Q(\bm{x}) \}^2  d\bm{x}}{\int Q^2(\bm{x})  d\bm{x}}.
	\end{equation}
Table \ref{tbl:l2_case1-1} shows $e_Q$ for a range of $\Delta$. It is clearly shown that gPC-LASSO outperforms the other two standard LS models. By using $h$-type refinement, i.e., reducing $\Delta$, $e_Q$ of LASSO and F-LASSO decreases at first. Then, when $\Delta$ is reduced from 2 to 1, $e_Q$ starts to grow. This result demonstrates a shortcoming of the $h$-type refinement of the LS inverse model, when only a limited number of observations is available. Refining the grid resolution of GRBF, the ratio of the number of unknown parameters to the number of observations of LASSO and F-LASSO increases from $N_k/N_o \simeq 1.4$ at $\Delta = 5$ to $N_k/N_o \simeq 51.4$ at $\Delta = 1$. Hence, the solution of the LS inverse model becomes more strongly dependent on the regularization as the grid is refined. It is worthwhile to note that gPC-LASSO also has a large number of unknown parameters; $N_k\times p^2 / N_o = 49$. However, due to the hierarchical nature of the basis functions shown in (\ref{eqn:non-negative_const}), the number of unknown parameters to estimate is effectively reduced to $N_k^0p^2$, in which $N_k^0$ is the number of non-zero elements in $\bm{\widehat{\beta}}^0$. In this example, it is found that $N^0_k$ is 8.
	
	In gPC-LASSO, although the grid space of $\mathcal{W}$ is bigger than the size of the true emission source, the estimated source surface is very close to the true surface. 
Since gPC-LASSO uses a modal method to approximate the sub-grid scale variations, gPC-LASSO can provide a good estimate of the source surface even when the local maximum of $Q(\bm{x})$ is located in the middle of the GRBF collocation points.
While F-LASSO correctly identifies the spatial pattern and the magnitude of the source (\ref{fig:comparison_LASSO} b), $Q^*(\bm{x})$ depends strongly on the choice of the collocation set $\mathcal{W}$. LASSO identifies the locations of large $Q(\bm{x})$, but fails to provide a good approximation of the function surface. The ratio of maximum value of the source surface, $\{\max Q^*(\bm{x})\} / \{\max Q(\bm{x}) \}$, is 0.99 and 0.93 for gPC-LASSO ($\Delta = 5$) and F-LASSO ($\Delta = 2$), respectively, while that of LASSO ($\Delta =2$) is 1.46, as LASSO tries to fit the data with a fewer number of stronger sources. 
	
	\begin{figure}
		\centering
		\includegraphics[width=0.8\textwidth]{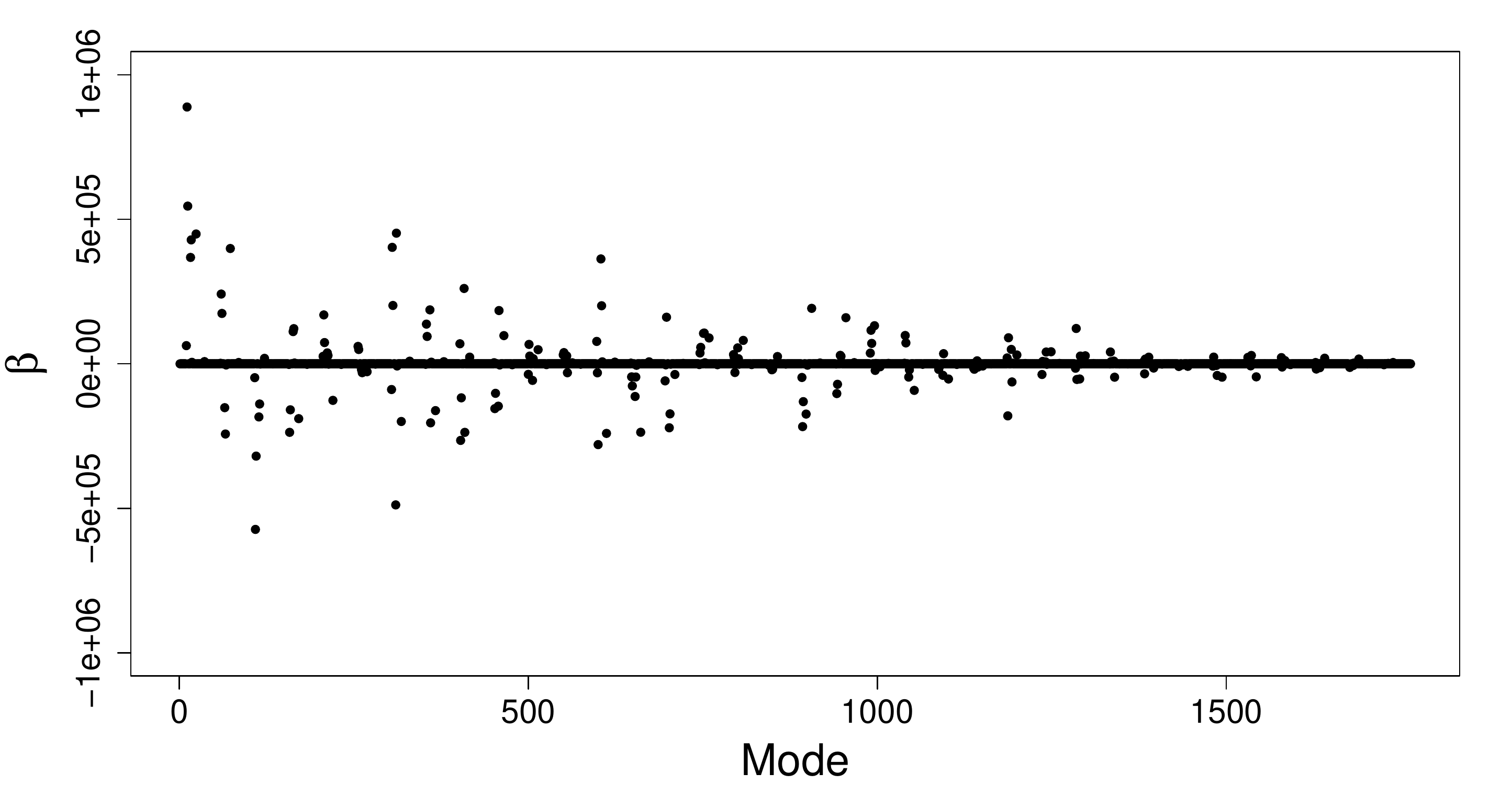}
		\caption{ The modal coefficients $\widehat{\bm{\beta}}$. }\label{fig:beta_mode}
	\end{figure}
	
Figure \ref{fig:beta_mode} shows the values of the modal coefficient $\widehat{\bm{\beta}}$. Among the total 1,764 parameters, there are only about 280 non-zero coefficients. It is shown that the magnitude of $\widehat{\bm{\beta}}$ decreases as the order of polynomial increases, i.e. going right on the horizontal axis. The magnitudes of  $\widehat{\bm{\beta}}$ of the highest order polynomials are very small, indicating that the maximum order of Legendre polynomial used in this case study is enough to resolve $Q(\bm{x})$. 
	
	\begin{figure}
		\centering
		\includegraphics[height=0.4\textwidth]{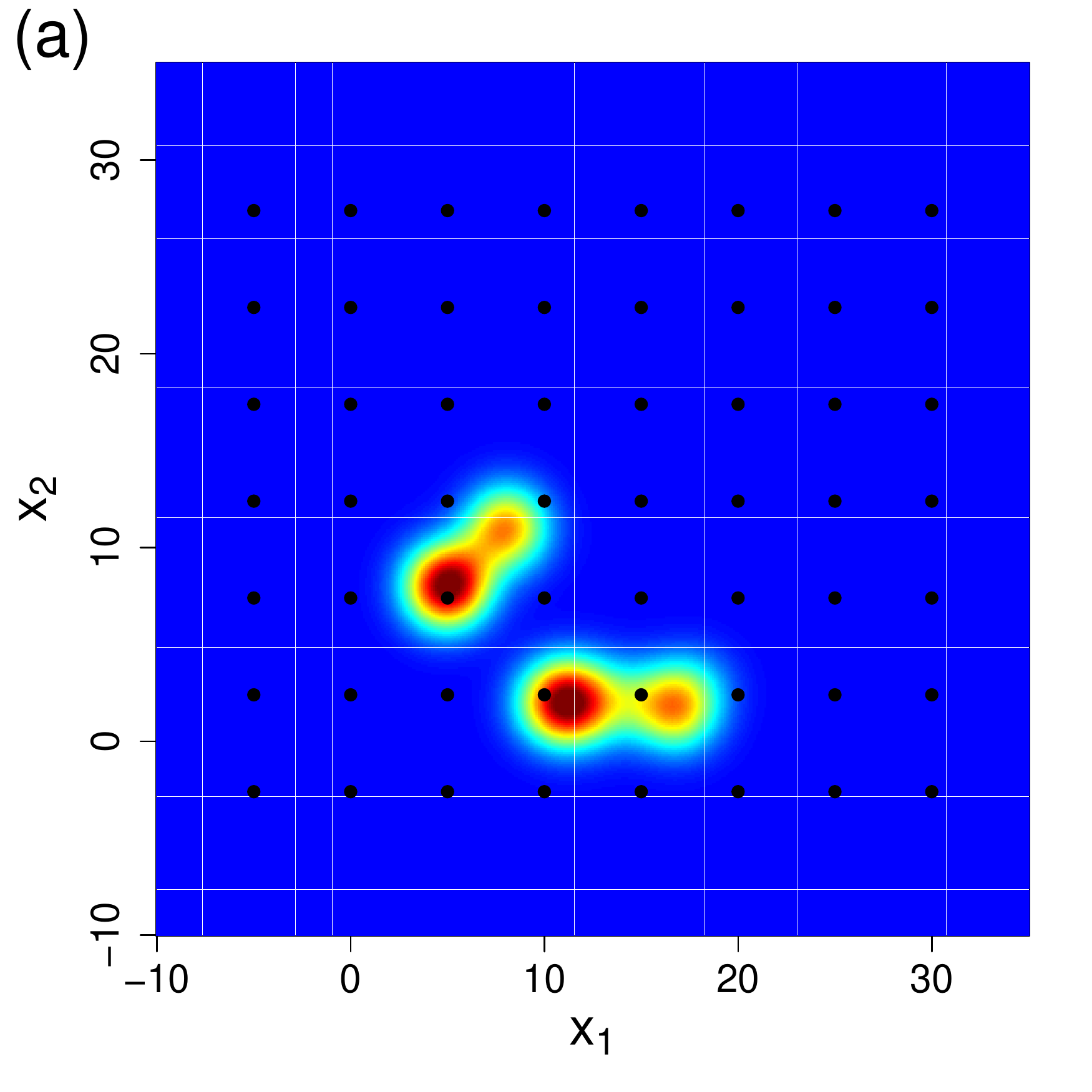}~
		\includegraphics[height=0.4\textwidth]{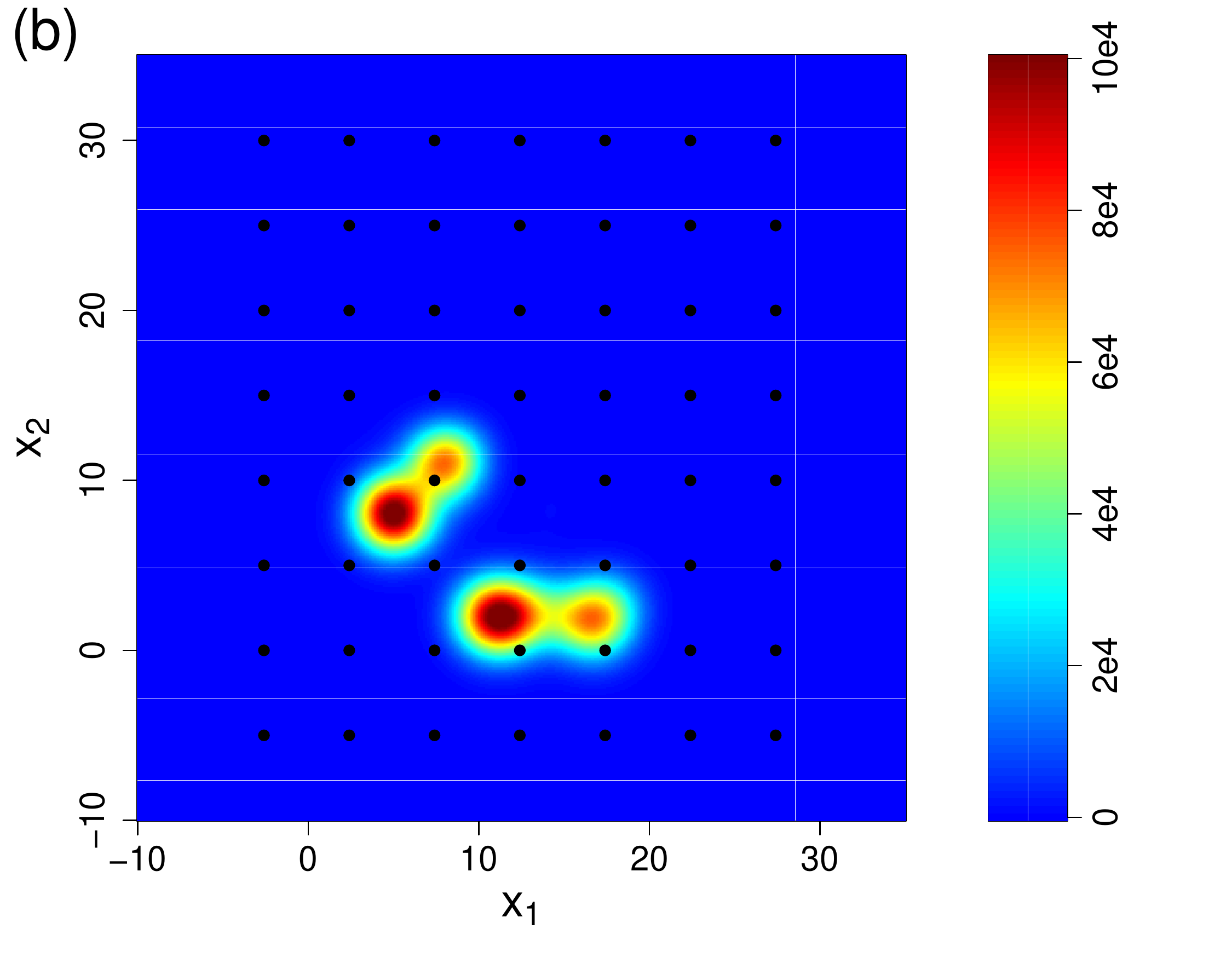}\\
		\caption{ The source surfaces from gPC-LASSO for two different collocation sets. }\label{fig:grid_shift}
	\end{figure}
	
	In figure \ref{fig:grid_shift}, gPC-LASSO is tested for two different collocation sets. From $\mathcal{W}$ used in figure \ref{fig:comparison_LASSO} (a), in figure \ref{fig:grid_shift} (a), $\mathcal{W}$ is shifted in the $x_1$-direction by $0.5 \Delta$ and, in  figure \ref{fig:grid_shift} (b), by $0.5 \Delta$ in the $x_2$-direction. As expected, it is shown that gPC-LASSO is not sensitive to the choice of the basis collocation set. The errors for the two new collocation sets are roughly the same, $e_Q \simeq 0.04$, which is very similar to the error of the reference case (figure \ref{fig:comparison_LASSO} a), $e_Q \simeq 0.03$.
	
	\begin{figure}
		\centering
		\includegraphics[height=0.4\textwidth]{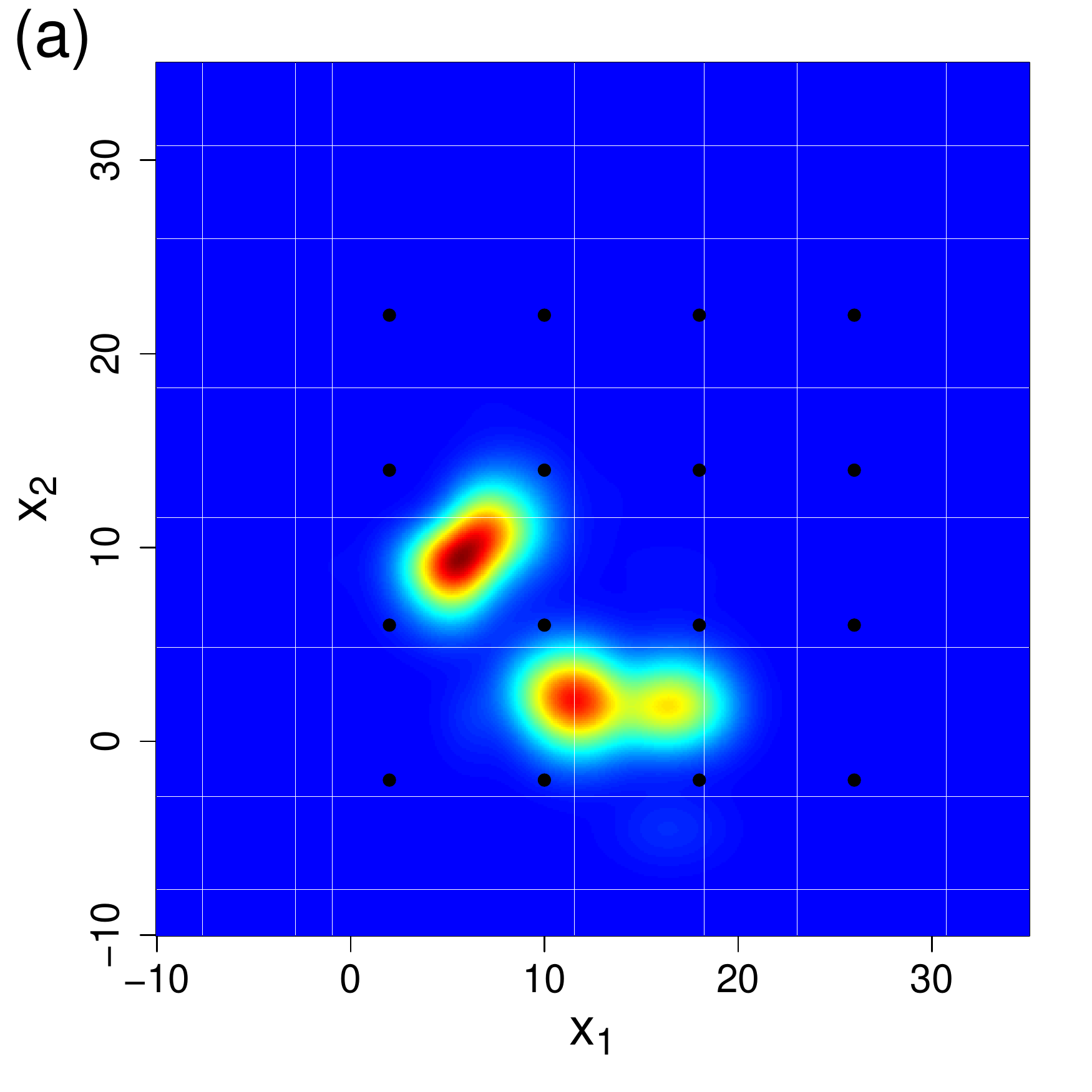}~
		\includegraphics[height=0.4\textwidth]{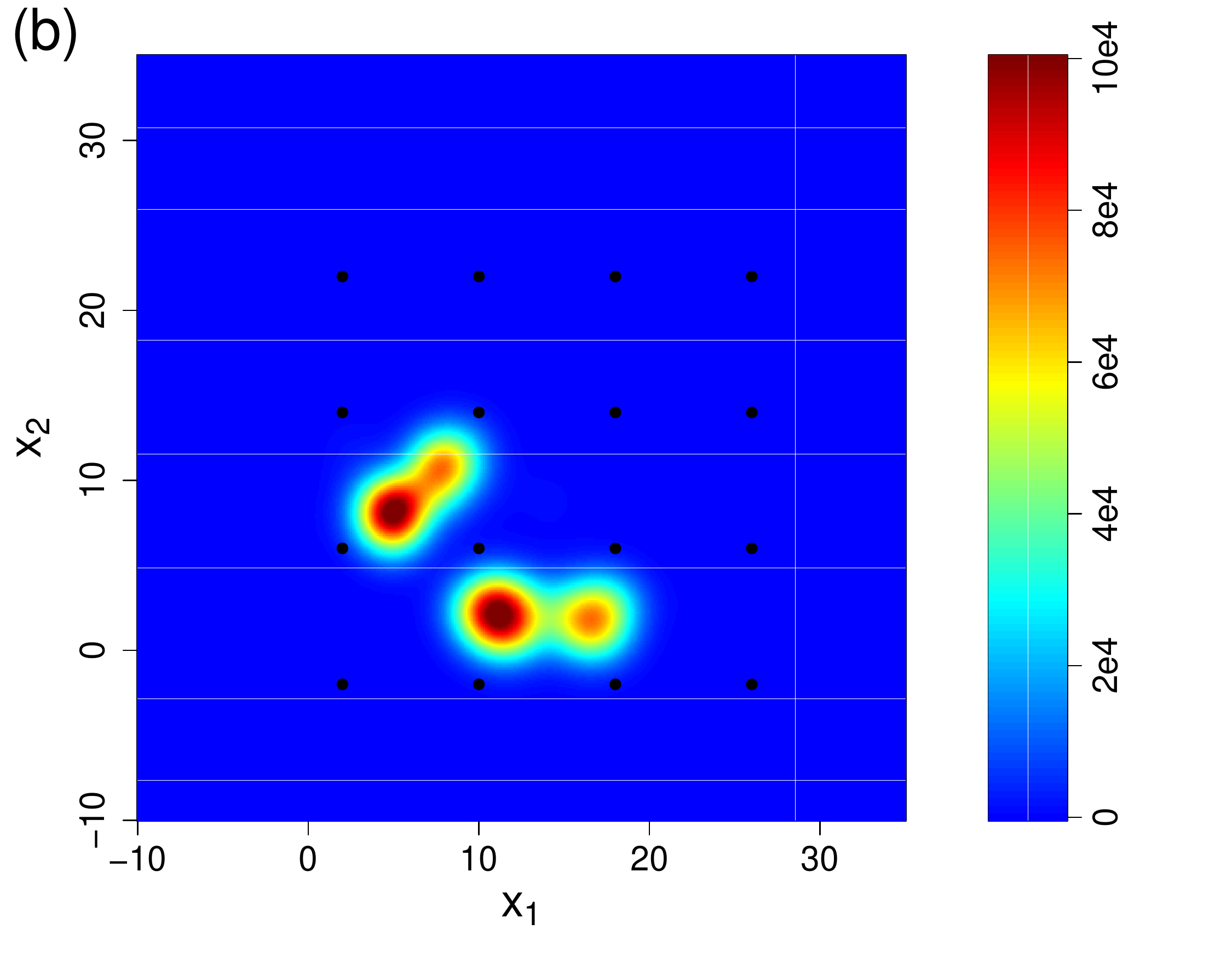}\\
		\caption{ The source surfaces from gPC-LASSO with the grid space $\Delta = 8$ and the maximum order of Legendre polynomial of (a) $P=4$ and (b) $P=8$. }\label{fig:case1_dx8}
	\end{figure}

\begin{table}
\center{
\caption{ Normalized $l_2$ error for a range of the maximum gPC order, $P$ ($\Delta = 8$).} 
\label{tbl:l2_case1-2}
\begin{tabular}{r|ccccc}
\hline \hline
$P$ & 2 & 4 & 6 & 8 & 10\\
\hline
$e_Q$ & 0.437 & 0.231 & 0.145 & 0.077 & 0.069\\
\hline \hline
\end{tabular}
}
\end{table}
	
Figure \ref{fig:case1_dx8} shows $Q^*(\bm{x})$ from gPC-LASSO at a coarser grid resolution, $\Delta = 8$. 
The same penalty parameters are used $\lambda_1 = 10^{-2}$ and $\lambda_2 = 10^{-6}$, while the scale parameter is changed, $c=0.15$. Figure \ref{fig:case1_dx8} (a, b) shows the estimated source surfaces for two different maximum order of gPC expansion; $P =4$ and 8. For $P=4$, gPC-LASSO under-resolves the source surface. When $P$ is increased to 8 (fig. \ref{fig:case1_dx8} b), even with the large grid space, gPC-LASSO is able to approximate $Q(\bm{x})$ fairly well. 

Table \ref{tbl:l2_case1-2} shows $e_Q$ as a function of $P$. This corresponds to the $p$-type refinement, where the grid resolution of GBRF is fixed and the maximum order of gPC expansion is increased to increase the fidelity of the estimation. Unlike the $h$-type refinement of the LS inverse model in Table \ref{tbl:l2_case1-1}, $e_Q$ monotonically decreases as $P$ increases in the range of $P$ tested. For a comparison, the ratio of the unknown parameters to the number of observations changes from $N_kp^2/N_o = 4$ at $P=2$ to $N_kp^2/N_o \simeq 53.8$ at $P = 10$.

\subsubsection{Effects of model parameters}
	
	\begin{figure}
		\centering
		\includegraphics[width=0.45\textwidth]{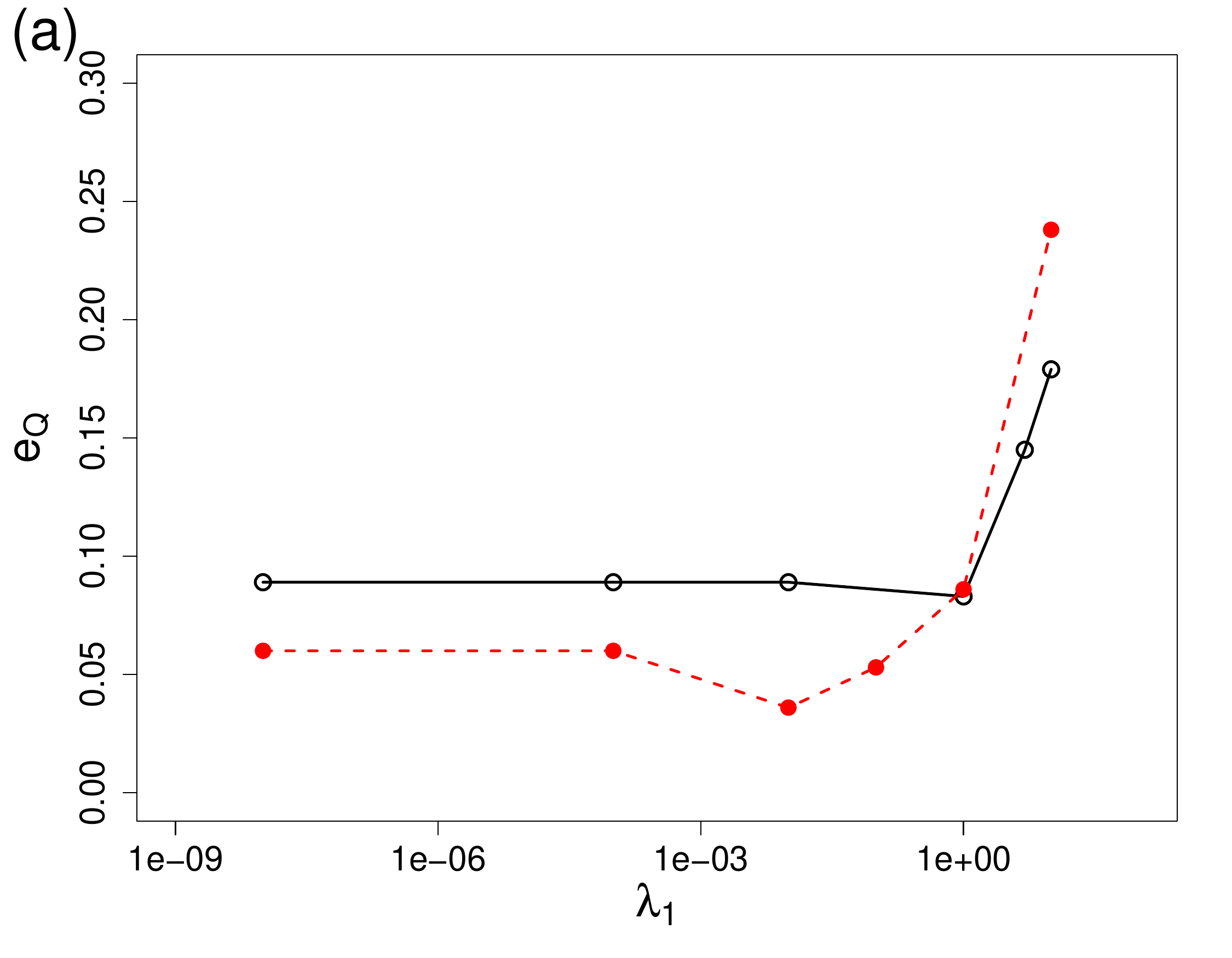}~
		\includegraphics[width=0.45\textwidth]{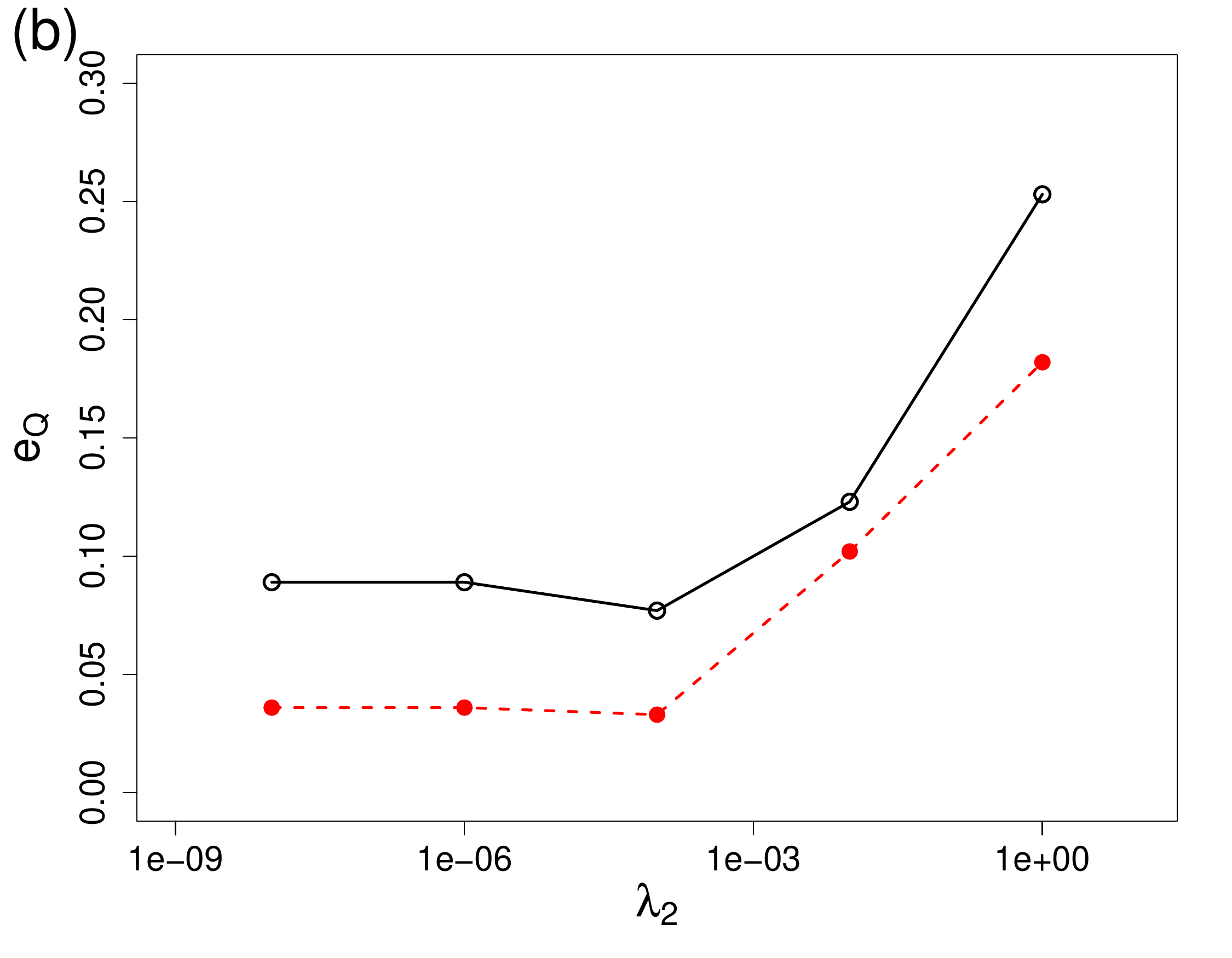}
		\caption{Changes of $e_Q$ with respect to the penalty parameters; (a) $\lambda_1$ and (b) $\lambda_2$. In (a), the solid symbol ($\bullet$) denotes $(\Delta = 5,P=5)$ and the hollow symbol ($\circ$) is for $(\Delta = 8,P=8)$.  }\label{fig:case1_err}
	\end{figure}
	
	In gPC-LASSO, there are two penalty parameters; $\lambda_1$ and $\lambda_2$. As discussed in section \ref{sec:optimization}, the $l_1$ penalty parameter $\lambda_1$ controls the sparsity of the mean $E[\bm{\beta}]$, while $\lambda_2$ is related with the spectral decay of the variance, $tr(Cov(\bm{\beta},\bm{\beta}))$. To show the effects of these penalty parameters on the solution, $e_Q$ is computed for a wide range of $\lambda_1$ ad $\lambda_2$. In figure \ref{fig:case1_err} (a), $e_Q$ is shown as a function of $\lambda_1$ for a fixed $\lambda_2 = 10^{-6}$. The model resolutions of gPC-LASSO are $(\Delta=5,P=5)$ and $(\Delta=8,P=8)$. In general, $e_Q$ is not very sensitive to $\lambda_1$ as long as $\lambda_1$ is sufficiently small $< 10^{-2}$. For $\Delta=5$, $e_Q$ seems to have a local minimum around $\lambda_1 = 10^{-2}$, which then increases rapidly for a larger $\lambda_1$. It is worthwhile to note that decreasing $\lambda_1$ does not have a significant effect on the solution, implying that the LASSO regularization does not play an important role in imposing sparsity in the solution. It is well known that, in a least-square regression problem, non-negativity constraint alone, without LASSO, is enough to recover sparsity \citep{Buckstein08,Slawski11,Wang11}. In gPC-LASSO, although non-negativity constraint is not directly imposed on $\widehat{\bm{\beta}}^0$, the linear constraint ($\mathcal{L}\widehat{\bm{\beta}} \ge 0$) effectively imposes the non-negativity constraint on the mean components,  $\widehat{\bm{\beta}}^0$, which explains why the generalized LASSO regularization for the mean component, $\| \bm{S}\widehat{\bm{\beta}}^0 \|_1$, does not have a significant effect on the solution.
	
	The effect of $\lambda_2$ is shown in figure \ref{fig:case1_err} (b). Again, $e_Q$ is not sensitive to $\lambda_2$.  However, when $\lambda_2$ becomes larger than $10^{-3}$, the higher-order modes of $\widehat{\bm{\beta}}$ are significantly suppressed, which results in the rapid increase of $e_Q$. 
	
	\begin{figure}
		\centering
		\includegraphics[width=0.7\textwidth]{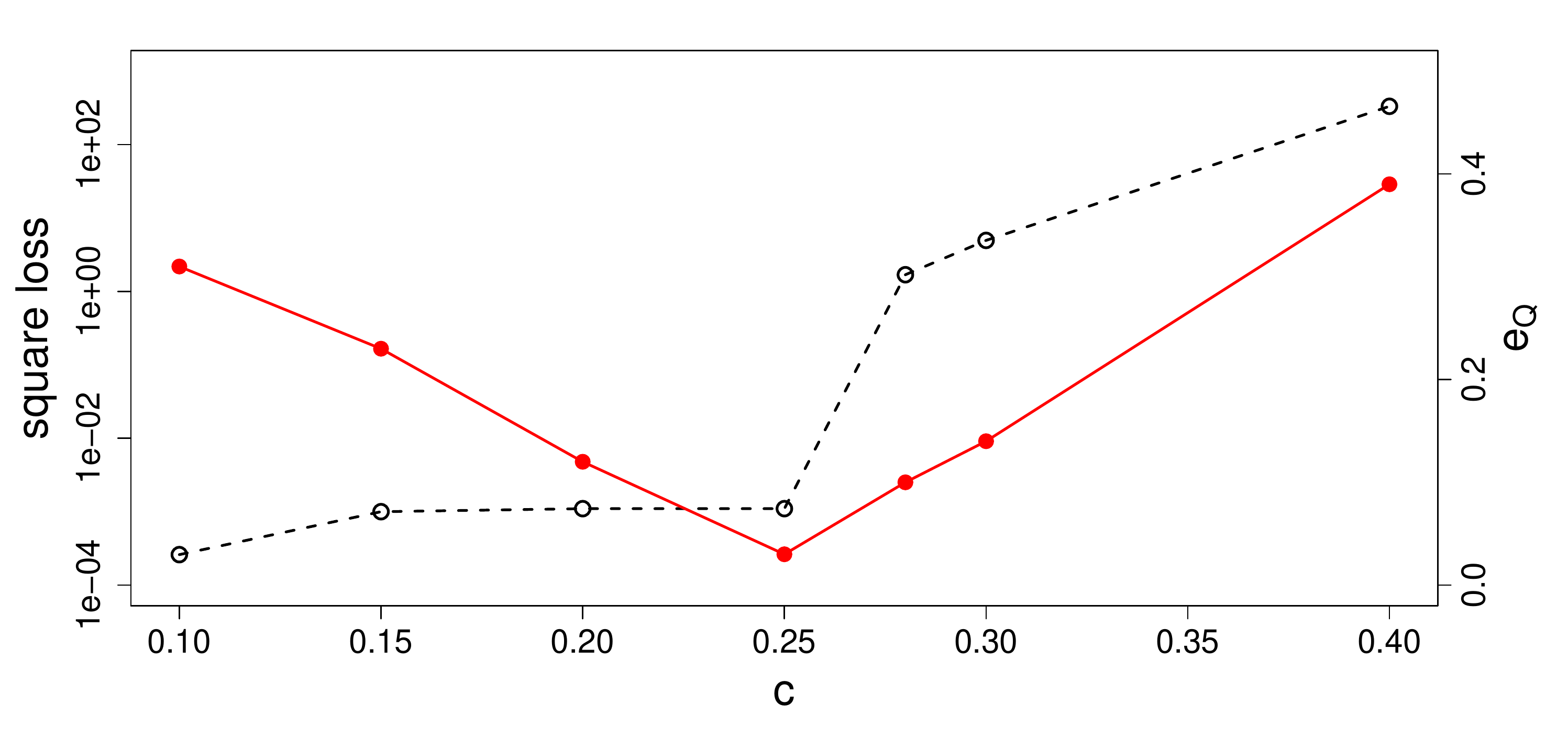}
		\caption{Effects of the scale parameter, $c$, on $Q^*(\bm{x})$. The hollow symbols ($\circ$) denote the square loss function (left axis) and the solid symbols ($\bullet$) are for $e_Q$ (right axis).}\label{fig:case_smoothness}
	\end{figure}
	
	Another important parameter of gPC-LASSO is the lengthscale of the basis Gaussian envelope, $c \Delta$. The scale parameter, $c$, essentially controls the smoothness of the estimated source surface. Figure \ref{fig:case_smoothness} shows the effects of $c$ on the square loss function,
	\[
	V(c) = \| \bm{\Phi} - \sum_{i=0}^M \widehat{\bm{X}}^i(c) \widehat{\bm{\beta}}^i  \|_2^2,
	\]
	together with the changes in $e_Q$. It is shown that $e_Q$ has a local minimum around $c = 0.25$, while $V(c)$ is almost constant up to $c=0.25$, which starts to increase rapidly afterward. As $c$ is changed from 0.25 to 0.28, $V(c)$ is increased by more than two orders of magnitude. For small $c$, gPC-LASSO tries to approximate $Q(\bm{x})$ with sharp, peaked polynomials, resulting in a highly oscillatory surface. Hence, although gPC-LASSO is able to find a solution to faithfully fit $\bm{\Phi}$,  $e_Q$ becomes large due to the high oscillation. On the other hand, for larger $c$, the basis polynomials become too smooth to approximate $Q(\bm{x})$, which makes both $V(c)$ and $e_Q$ grow. 
The optimal choice of $c$ seems to be related with both $\Delta$ and the lengthscale of $Q(\bm{x})$. It is challenging to decide the optimal $c$ a priori.
However, figure \ref{fig:case_smoothness} provides a guidance on how to select the scale parameter by performing a set of numerical tests, e.g., find the maximum $c$ which satisfies $V(c) < \delta_c$ for a threshold level $\delta_c$.

\subsubsection{Noisy observations}

\begin{figure}
  \centering
    \includegraphics[width=0.45\textwidth]{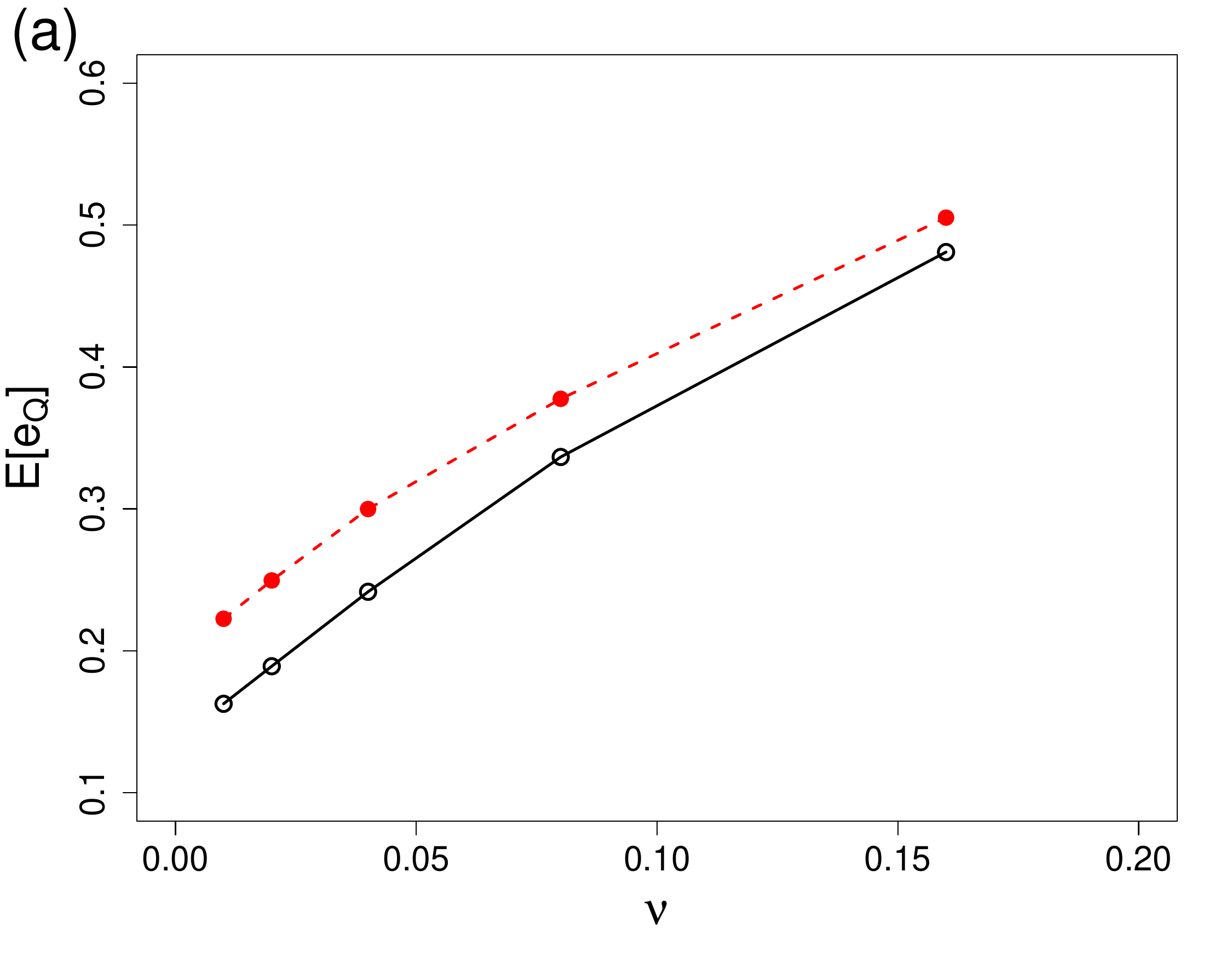}~
    \includegraphics[width=0.45\textwidth]{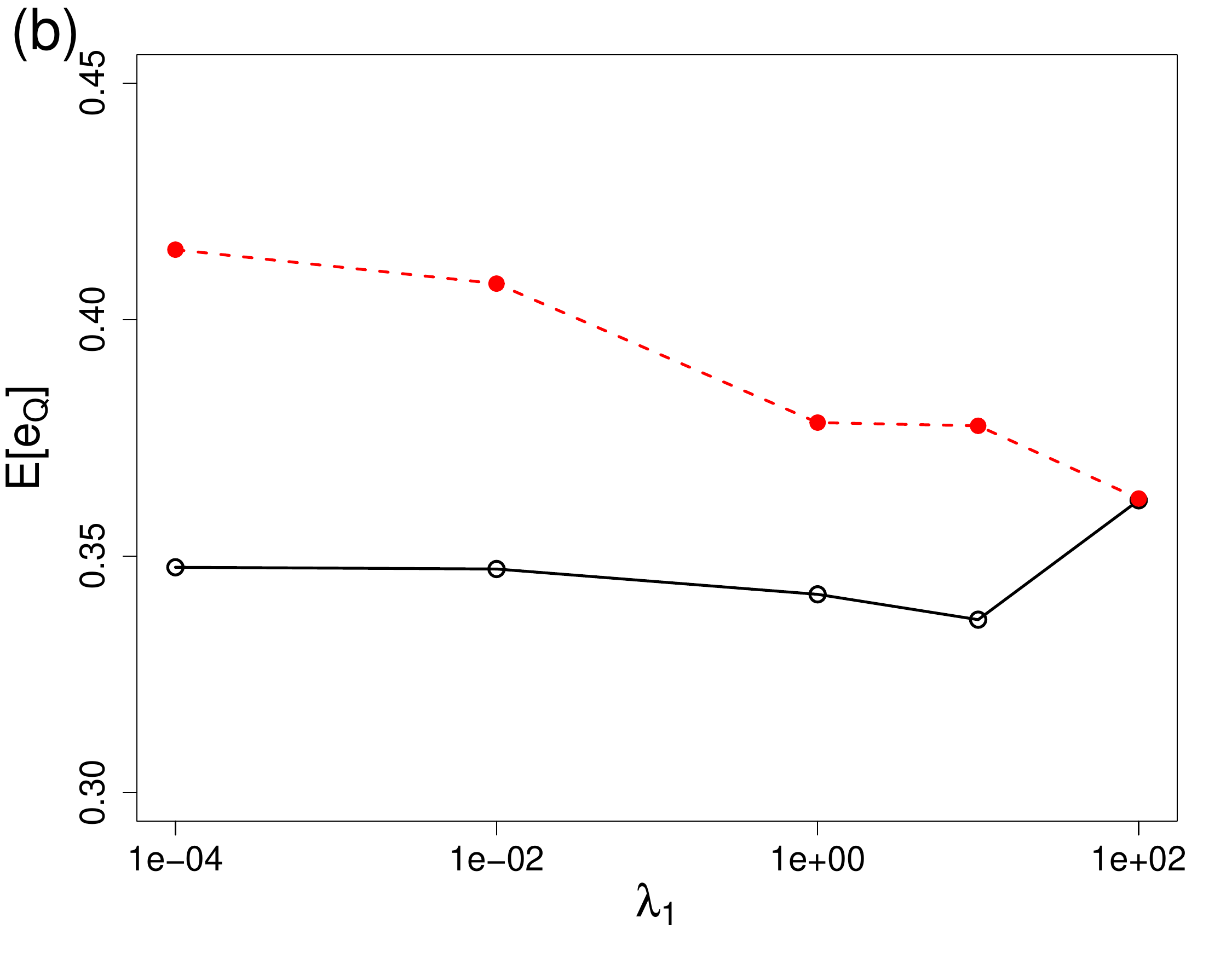}
    \caption{ Effects of (a) noise level ($\nu$) and (b) $l_1$-penalty parameter ($\lambda_1$) on the estimation error. The solid symbol ({\color{red}$\bullet$}) denotes $(\Delta = 5,P=5)$ and the hollow symbol ($\circ$) is for $(\Delta = 8,P=8)$ } \label{fig:eff_noise_d3}
\end{figure}

To show the effects of the observation error on gPC-LASSO, the 36 observations are perturbed by an additive noise;
	\begin{equation} \label{eqn:noise}
	\bm{\Phi}^* = \max (\bm{\Phi}+ \bm{\epsilon},\bm{0}),
	\end{equation}
in which $\epsilon$ is a Gaussian random variable,
\[
\bm{\epsilon} \sim \mathcal{N}(\bm{0}, \sigma^2 \bm{I}),~~\text{and}~~\sigma = \frac{\nu}{N_o} \| \bm{\Phi} \|_1.
\]
The parameter, $\nu$, decides the signal-to-noise ratio. 
For a quantitative comparison, an ensemble error is computed from a Monte Carlo simulation with 200 samples;
	\[
	E_{\bm{\epsilon}}[e_Q] \simeq \frac{1}{200}\sum_{i=1}^{200} e_Q(\bm{\epsilon}_i).
	\]
Figure \ref{fig:eff_noise_d3} shows the effects of the noise on the estimation error. Two resolutions are used for the comparison, $(\Delta=5,P=5)$ and $(\Delta=8,P=8)$.

In figure \ref{fig:eff_noise_d3} (a), the effects of the noise level are shown. The penalty parameters are fixed at $\lambda_1 = 10$, $\lambda_2 = 0.01$, and $\gamma = 0.25$. Because we consider the problem of estimating the source surface from a small number of observations ($N_o = 36$), it is not surprising to see that the inverse model is sensitive to the noise. For $\Delta = 5$, at $\nu = 0.01$, the ensemble error is about 0.22. while that of $\Delta = 8$ is 0.16. It is found that using a low resolution GRBF with a higher-order gPC mode makes gPC-LASSO less susceptible to the noise in the data. 

The effect of LASSO for the noisy observation is shown in figure \ref{fig:eff_noise_d3} (b). In this set of experiments, the noise level and the other penalty parameters are fixed at $\nu = 0.08$, $\lambda_2 = 0.01$, and $\gamma = 0.25$. It is shown that, for $\Delta = 8$, the $l_1$-penalty does not play an important role. Similar to the noiseless case (figure \ref{fig:case1_err}), the ensemble error is insensitive to $\lambda_1$ for smaller values of $\lambda_1$, and starts to grow when $\lambda_1 \ge 10$. On the other hand, for a finer resolution ($\Delta =5$), the $l_1$-penalty makes gPC-LASSO more resistant to the noise. The ensemble error is shown to be a monotonically decreasing function of $\lambda_1$ for $10^{-4} \le \lambda_1 \le 10^2$.
	
%
	
	\subsection{Case study 2}
	
	\begin{figure}
		\centering
		\includegraphics[height=0.4\textwidth]{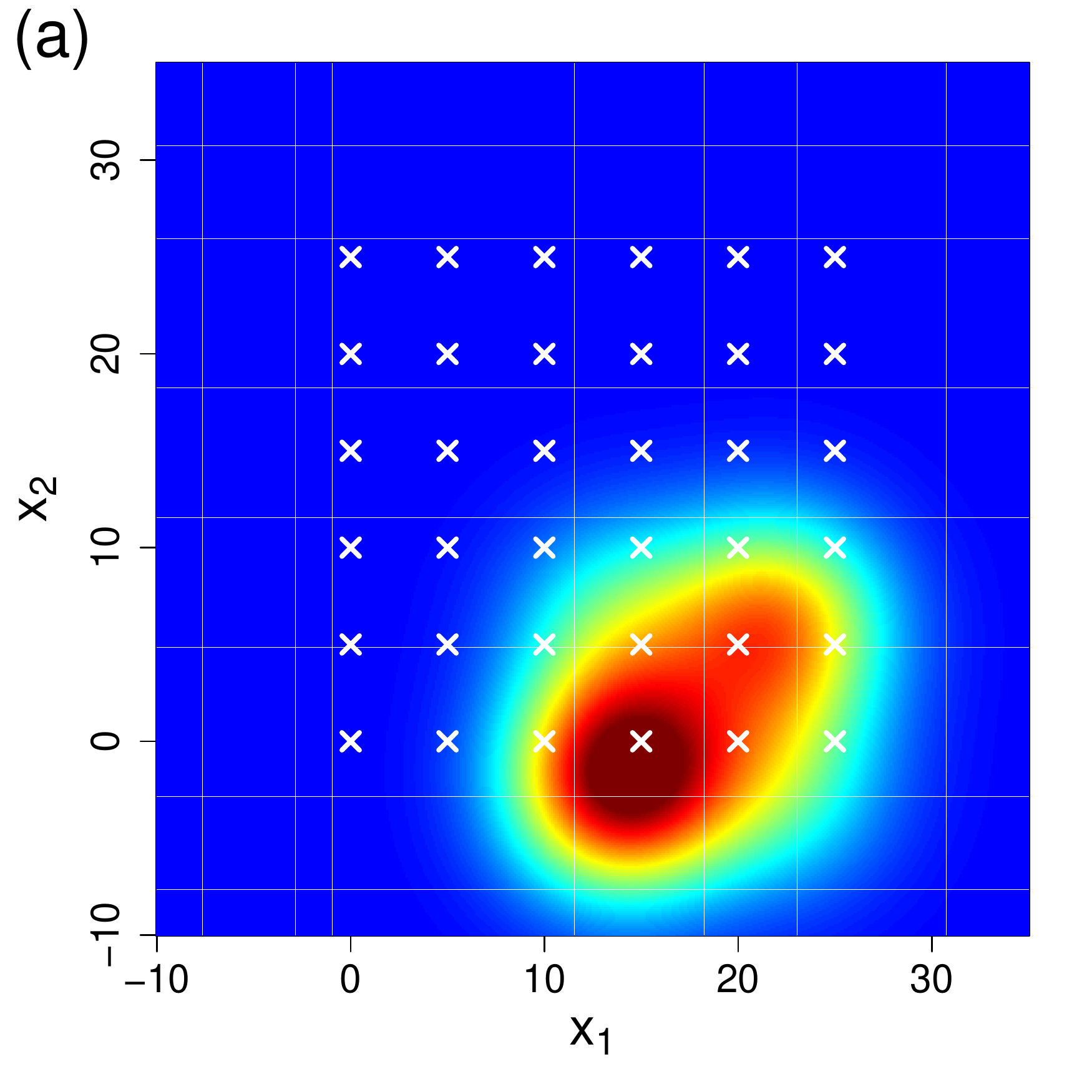}~
		\includegraphics[height=0.4\textwidth]{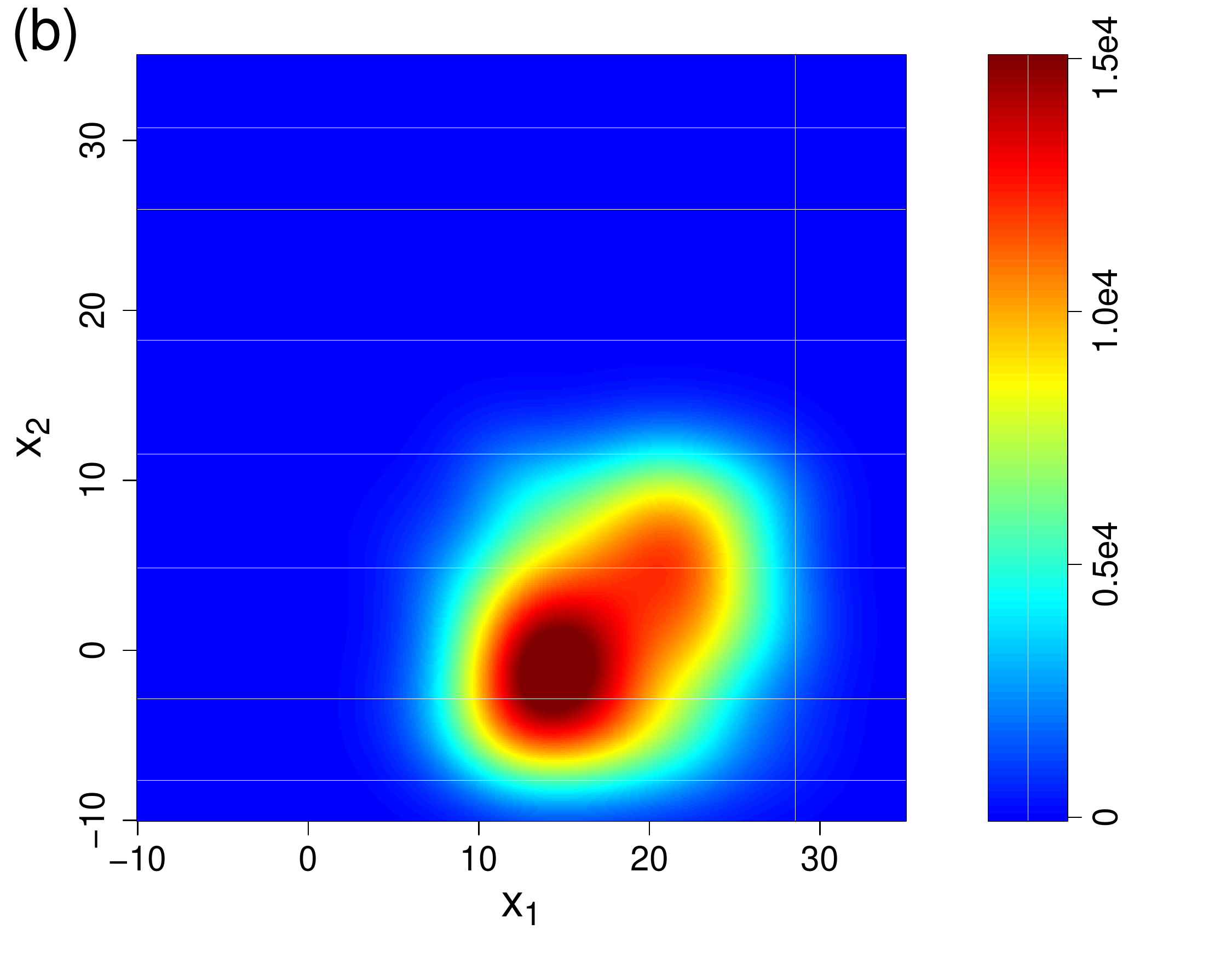}\\
		\includegraphics[height=0.4\textwidth]{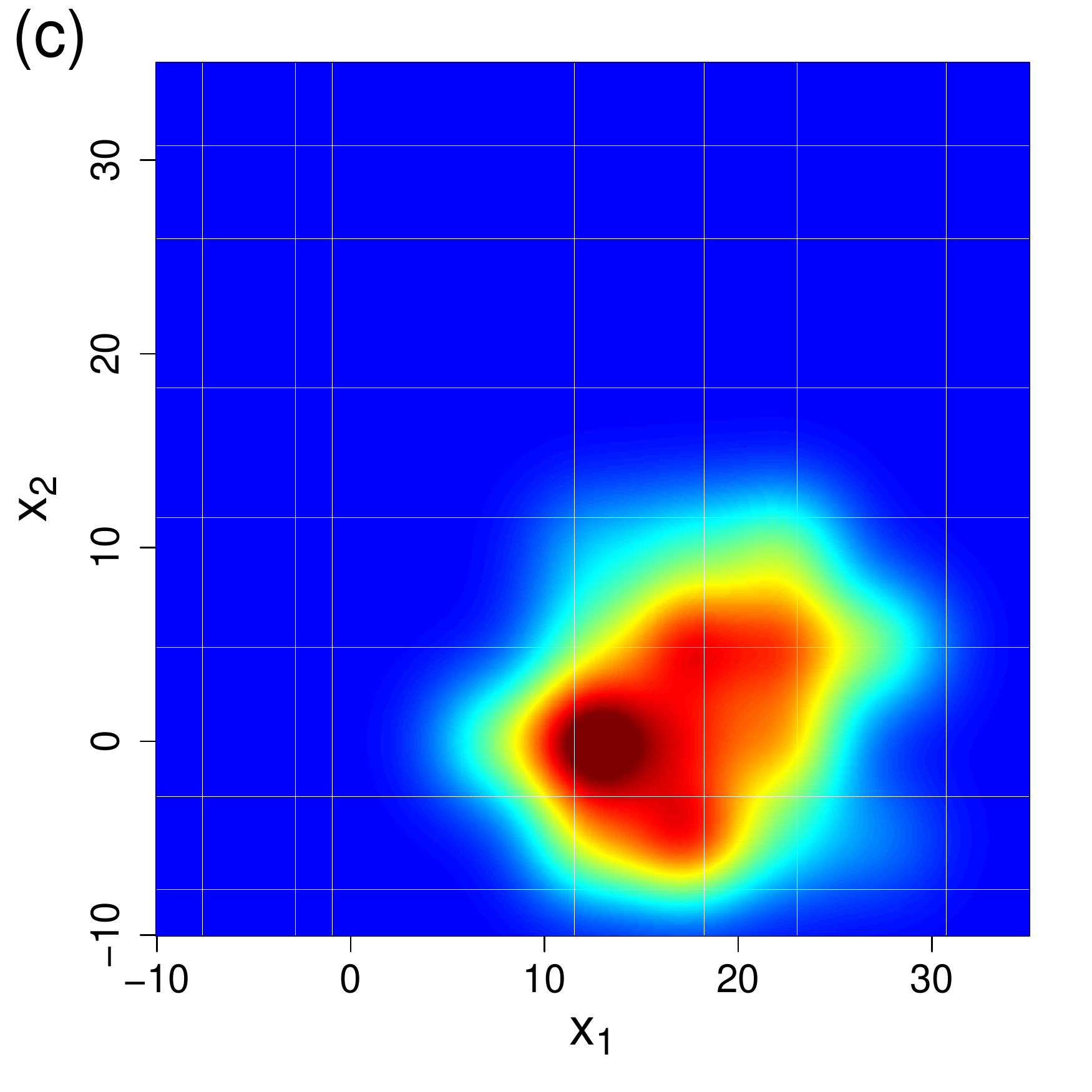}~
		\includegraphics[height=0.4\textwidth]{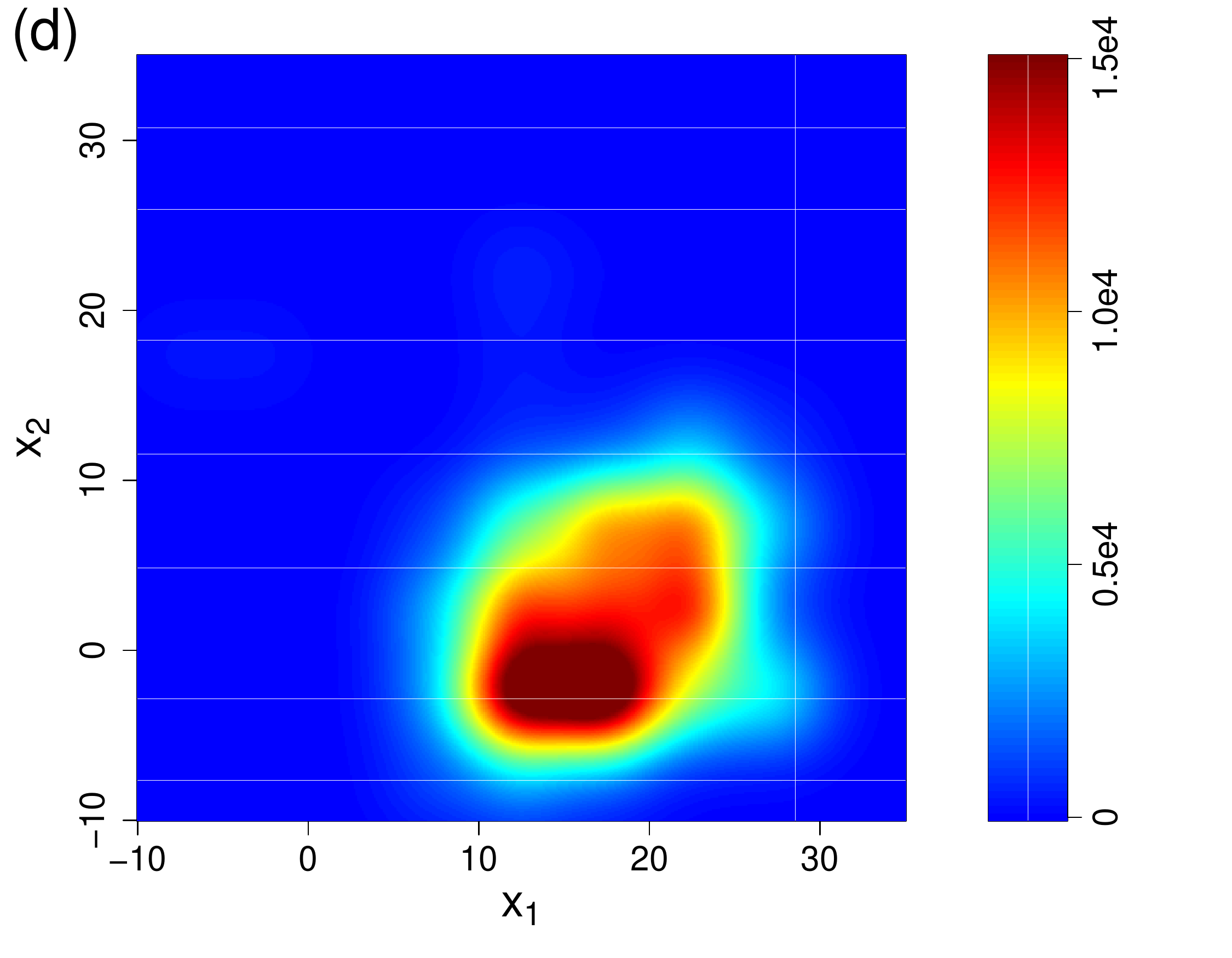}
		\caption{ (a) True source surface and the estimated source surfaces by (b) gPC-LASSO, (c) LASSO, and (d) fused LASSO. The white crosses in (a) indicate the locations of the sensors.  }\label{fig:case2}
	\end{figure}
	

\begin{table}
\center{
\caption{ Normalized $l_2$ errors for $\Delta = 5$.} 
\label{tbl:l2_case2-1}
\begin{tabular}{r|ccc}
\hline \hline
& gPC-LASSO & F-LASSO & LASSO\\
\hline
$e_Q$ & 0.04 & 0.13 & 0.14\\
\hline \hline
\end{tabular}
}
\end{table}

	In the second case study, the emission surface, $Q(\bm{x})$, has a larger lengthscale than the grid space $\Delta$ (figure \ref{fig:case2} a), and the emission surface extends to the area not covered by the sensors. In figure \ref{fig:case2} (b--d), $Q^*(\bm{x})$ from gPC-LASSO is compared with the solutions of F-LASSO and LASSO. The parameters of gPC-LASSO are chosen the same with the first case study, $\Delta = 5$, $P=5$, $\lambda_1 = 10^{-2}$, and $\lambda_2 = 10^{-6}$, except for the scaling parameter $c=0.5$.  For LASSO and F-LASSO, the model parameters are $\Delta =5$, $c=0.5$, $\lambda_1 = 10^{-2}$, and $\lambda_2 = 10^{-6}$. It is again shown that gPC-LASSO provides a better approximation of $Q(\bm{x})$. For a quantitative comparison, the normalized $l_2$-error is listed in Table \ref{tbl:l2_case2-1}.
	
	\begin{figure}
		\centering
		\includegraphics[width=0.6\textwidth]{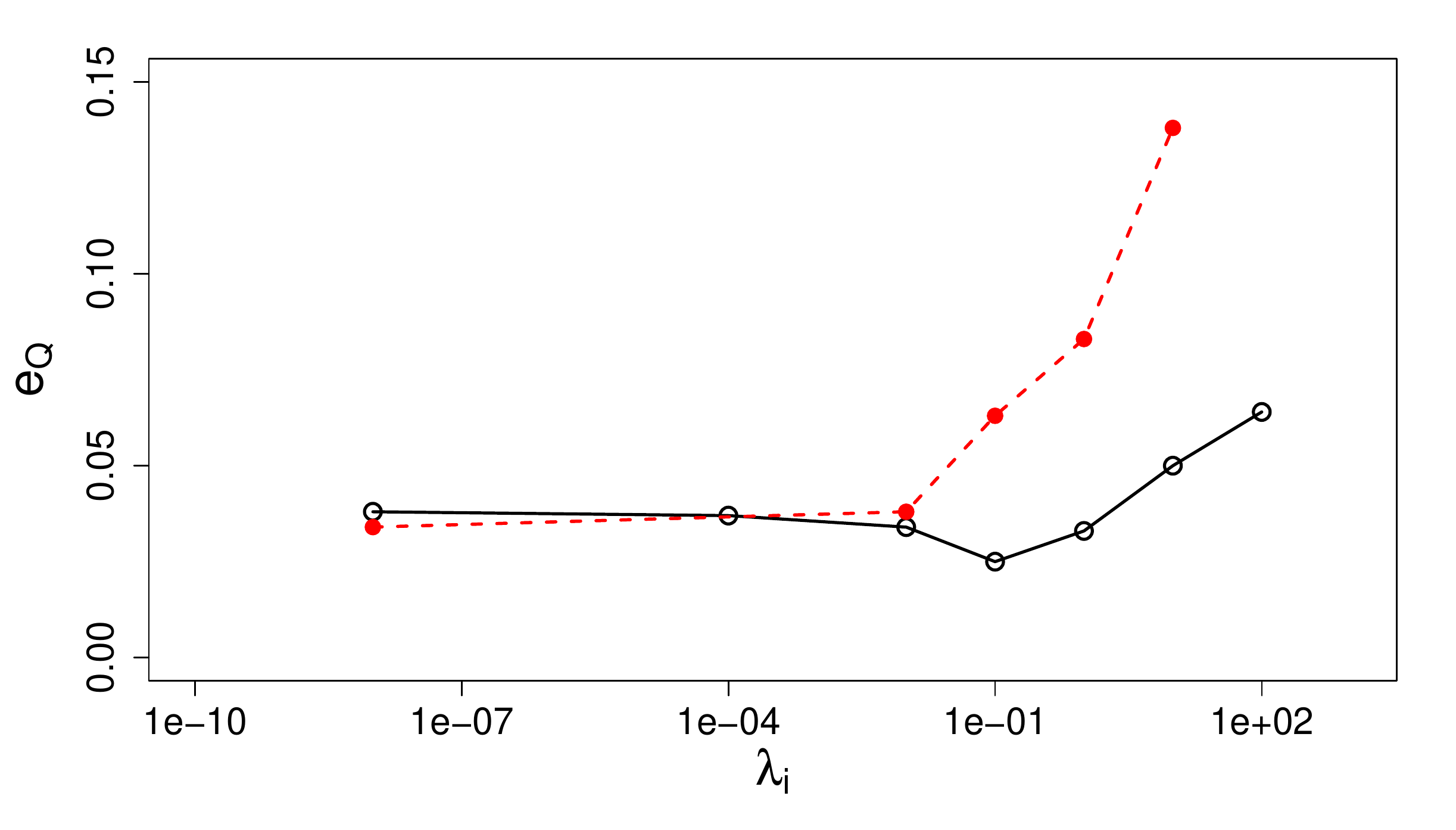}
		\caption{ Dependence of $e_Q$ on the penalty parameters $\lambda_1$ ($\bullet$) and $\lambda_2$ ($\circ$).} \label{fig:case2_err}
	\end{figure}
	
	Figure \ref{fig:case2_err} shows the behavior of $e_Q$ with respect to the penalty parameters, $\lambda_1$ and $\lambda_2$. Similar to the previous results (figure \ref{fig:case1_err}), it is shown that $e_Q$ is not sensitive to $\lambda_1$ and $\lambda_2$. Although the normalized $l_2$ error, $e_Q$, shows a local minimum around $\lambda_2 = 10^{-1}$, the difference between the local minimum and $e_Q$ at smaller $\lambda_2$ is only 0.01.
	
%

\begin{figure}
  \centering
    \includegraphics[width=0.45\textwidth]{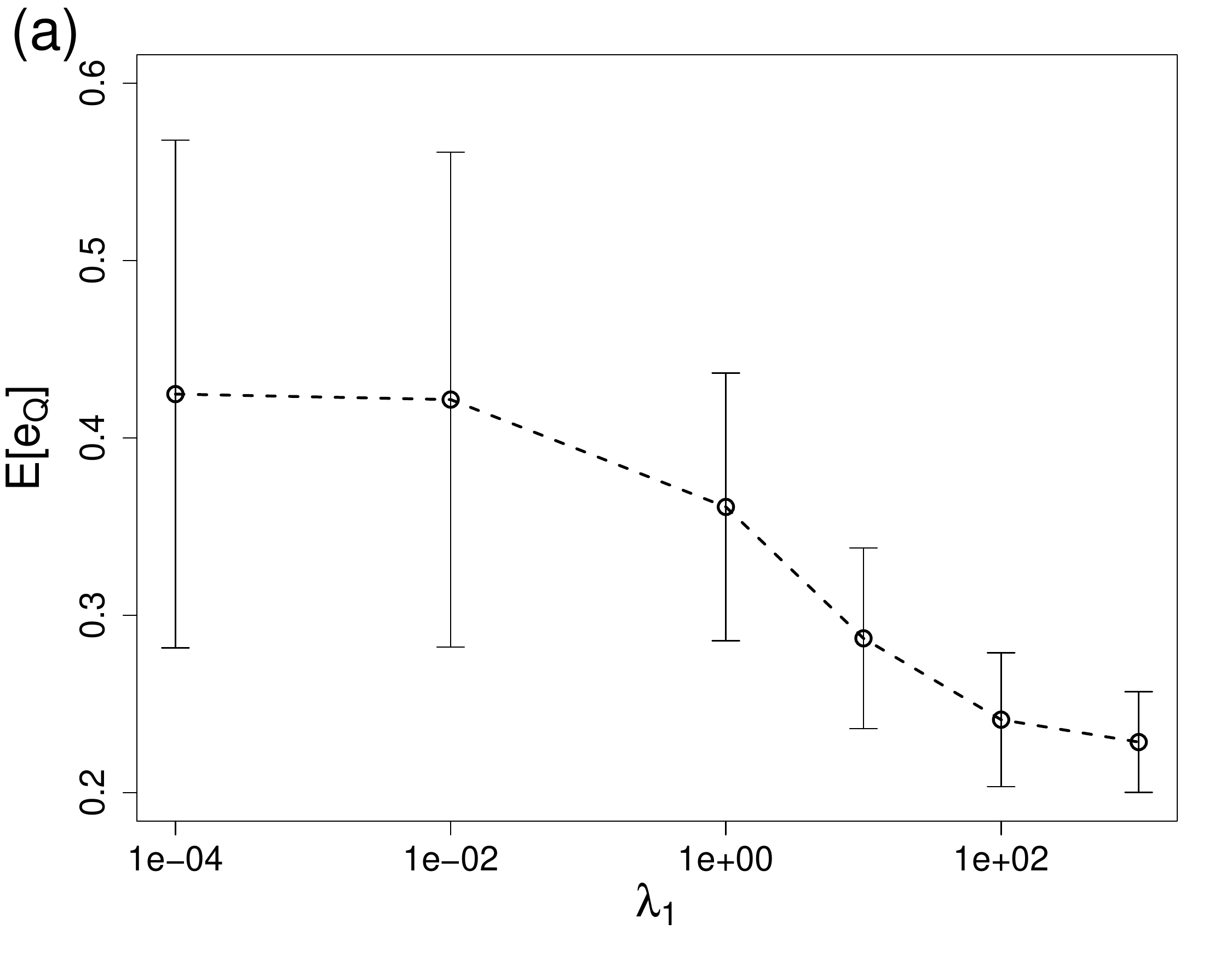}~
    \includegraphics[width=0.45\textwidth]{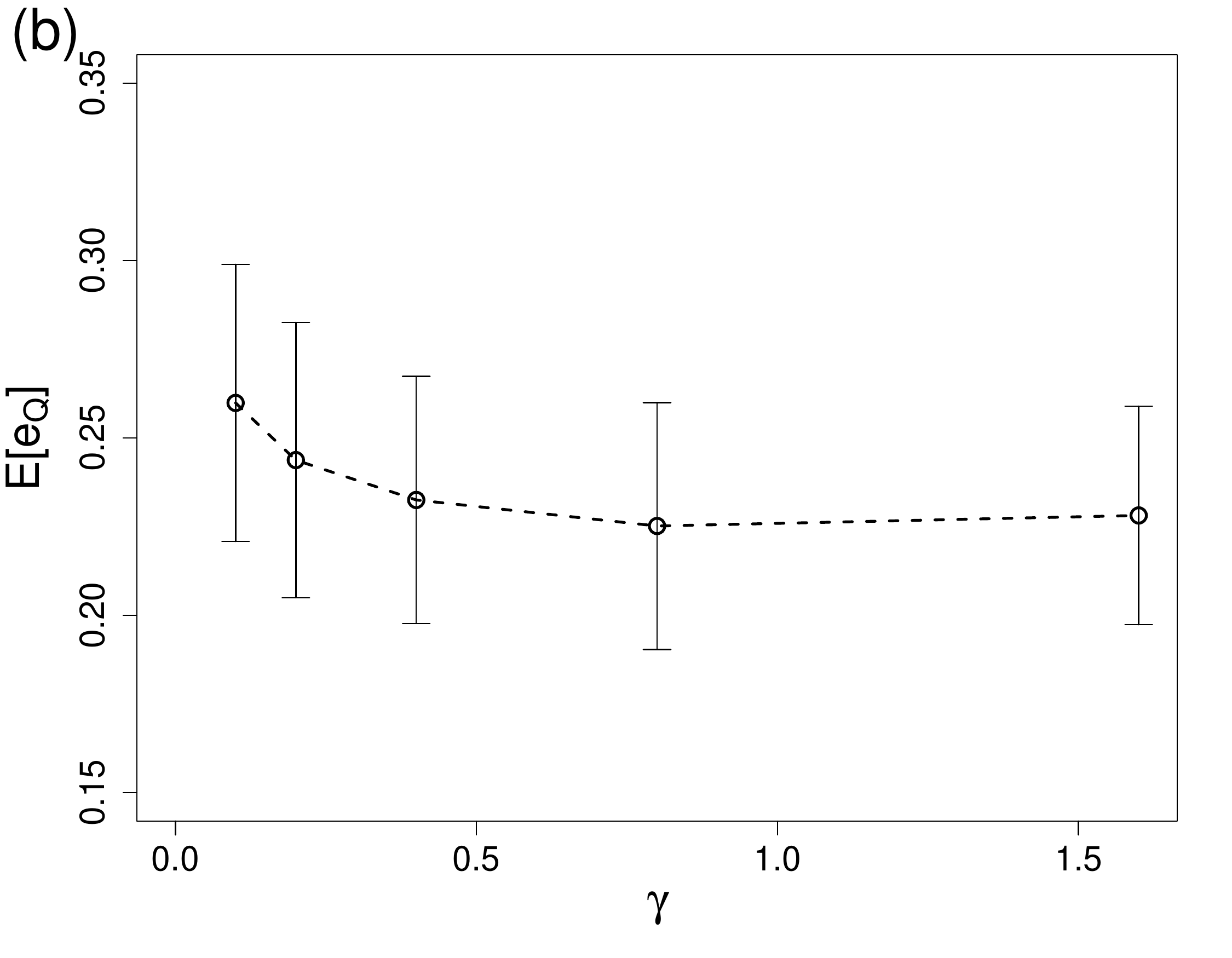}
    \caption{ Effects of (a) $l_1$-penalty parameter ($\lambda_1$) and (b) fussed-LASSO parameter ($\gamma$) on the estimation error.  The error bars denote one standard deviation of $e_Q$. } \label{fig:eff_noise_d8}
\end{figure}

Figure \ref{fig:eff_noise_d8} (a) shows the effects of the $l_1$-penalty parameter ($\lambda_1$) on $Q^*(\bm{x})$ for the noisy observation, (\ref{eqn:noise}). The noise level is $\nu = 0.08$ and $\lambda_2$ is fixed, $\lambda_2=10^{-2}$. The expectation and standard deviation of $e_Q$ is computed by a Monte Carlo simulation with 200 samples. For the noisy observations, it is clearly shown that the $l_1$-penalty has a significant impact on the accuracy of gPC-LASSO. When $\lambda_1$ is very small, gPC-LASSO finds a solution which overfits the noisy data. In other words, gPC-LASSO results in a solution with a highly oscillatory surface to minimize the square loss function, $\| \bm{\Phi} - \bm{X}\bm{\beta}\|^2_2$, for the noisy observation. For a larger $\lambda_1$, the smoothness of the solution is recovered by the generalized LASSO regularization, which makes $E_{\bm{\epsilon}}[e_Q]$ smaller. It is shown that, as $\lambda_1$ increases, the standard deviation of $e_Q$ is also reduced, implying that the generalized LASSO regularization also makes the estimation more robust to the noise.

The effects of the fused-LASSO parameter, $\gamma$, are shown in figure \ref{fig:eff_noise_d8} (b). For this test, $\lambda_1 = 100$ and $\nu = 0.08$ are used. It is shown that using a higher value of $\gamma$ results in a smaller error. However, the effects of $\gamma$ on $E_{\bm{\epsilon}}[e_Q]$ are not as significant as $\lambda_1$. There is about 10\% reduction in the error as $\gamma$ is increased from 0.1 to 0.8.
	
	\section{Summary}  \label{sec:summary}
In this study, we present a $hp$-inverse model to estimate a source function from a limited number of data for an advection-diffusion problem. 
One of the standard methods of approximating a smooth source function, $Q(\bm{x}$), is to discretize the computational domain by a mesh system ($\mathcal{W}$) and compute the coefficients, $\bm{\beta}(\mathcal{W})$, of a basis function, such as GRBF. However, in such a mesh-based inverse model, the estimated function surface, $Q^*(\bm{x})$, strongly depends on the choice of $\mathcal{W}$. To remove the dependence on the fixed mesh system, we formulate a stochastic least-square inverse model on a random mesh system, $\mathcal{W}^*(\omega)$. The generalized polynomial chaos expansion (gPC) is employed to approximate the resulting stochastic functions; the source-receptor relation $\bm{X}(\omega)$ and the source strength $\bm{\beta}(\omega)$. 

By using gPC, a $hp$-inverse model is formulated, where $Q(\bm{x})$ is approximated by hierarchical polynomials.
The $hp$-refinement approach has advantages over the conventional mesh-based method in that $Q^*(\bm{x})$ is not as strongly dependent on $\mathcal{W}$, and the spatial sparsity in $Q(\bm{x})$ can be more effectively recovered. The non-negativity constraint of $Q(\bm{x}) \ge 0,~\forall \bm{x} \in D$, is replaced by a linear constraint, $\mathcal{L} \widehat{\bm{\beta}} \ge \bm{0}$, by comparing the modal coefficients of gPC with the nodal coefficients of a stochastic collocation method. Finally, a mixed $l_1$ and $l_2$ regularization is proposed based on the hierarchical nature of the basis polynomials. An ADMM algorithm is presented to solve the regularized optimization problem.
	
The solution behavior of the proposed method (gPC-LASSO) is investigated for two case studies and the model error is compared with the mesh-based least square inverse methods with two most widely used regularization methods, LASSO and fused LASSO. It is confirmed that gPC-LASSO is not sensitive to the choice of $\mathcal{W}$ and provides a very good approximation to the source surface even when the number of unknown parameters is more than 40 times larger than the number of data. It is shown that gPC-LASSO outperforms both LASSO and fused LASSO. For the noise-free data, the regularization does not play an important role for gPC-LASSO because the non-negativity constraint alone is enough to explore the sparsity in the solution. However, when noisy is added to the observations, the regularization provides a more robust approximation by enforcing the smoothness in the solution.
	

In summary, we show that a $hp$-inverse model can be developed by converting the deterministic problem to a stochastic problem and the $hp$-refinement capability has an advantage in exploring the sparsity structure in the data. We expect that the proposed framework can be applied to a broader class of problems, such as a general data-driven function estimation problem for a smooth, non-negative function. It should be noted that we limited our focus on a two-dimensional source inverse model for a steady emission in this study. It is a subject of the follow-up study how to generalize the framework to consider a much more complex problem of nonlinear inversion or estimation of unsteady emission sources.
	


	\bibliographystyle{model4-names}
	\bibliography{ref_inv}

\end{document}